\documentclass[12pt]{article}

\usepackage{scicite}

\usepackage{times}
 \usepackage{tabularx}
 \usepackage{graphicx}
 \usepackage{caption}
 \usepackage{float}
 \usepackage{xcolor}
 \usepackage{calrsfs}
\DeclareMathAlphabet{\pazocal}{OMS}{zplm}{m}{n}

\newcommand{\Lb}{\pazocal{L}}
\newcommand\farcs{\mbox{$.\!\!^{\prime\prime}$}}%

\newenvironment{sciabstract}{%
\begin{quote} \bf}
{\end{quote}}

\title{Direct Imaging and Astrometric Detection of a Gas Giant Planet Orbiting an Accelerating Star}
\author{Thayne Currie$^{1,2,3}$\footnote{To whom correspondence should be addressed; E-mail:  currie@naoj.org.}, G. Mirek Brandt$^{4}$, Timothy D. Brandt$^{4}$, Brianna Lacy$^{5,6}$,\\
Adam Burrows$^{5}$, Olivier Guyon$^{1,7,8}$, 
Motohide Tamura$^{7,9,10}$, 
Ranger Y. Liu$^{11}$\footnote{Present address: Parsons School of Design, New York, NY 10011, USA}, \\Sabina Sagynbayeva$^{12}$, Taylor Tobin$^{13}$\footnote{Present address: Department of Astronomy, University of Michigan, Ann Arbor, MI 48109, USA}, Jeffrey Chilcote$^{13}$, Tyler Groff$^{14}$, \\
Christian Marois$^{15,16}$, William Thompson$^{16}$, Simon J. Murphy$^{17}$\footnote{Present address: Centre for Astrophysics, University of Southern Queensland, Toowoomba, QLD 4350, Australia}, Masayuki Kuzuhara$^{7,9}$,\\ Kellen Lawson$^{14}$, Julien Lozi$^{1}$, Vincent Deo$^{1}$, Sebastien Vievard$^{1}$, Nour Skaf$^{1}$,\\ Taichi Uyama$^{9,18}$,
Nemanja Jovanovic$^{19}$, Frantz Martinache$^{20}$, N. Jeremy Kasdin$^{21}$,\\Tomoyuki Kudo$^{1}$, Michael McElwain$^{14}$,
Markus Janson$^{22}$, John Wisniewski$^{23}$, \\
Klaus Hodapp$^{24}$, Jun Nishikawa$^{9,10}$, Krzysztof Hełminiak$^{25}$, Jungmi Kwon$^{10}$, \\Masa Hayashi$^{9}$
\\
\vspace{-0.025in}
\small{$^{1}$Subaru Telescope, National Astronomical Observatory of Japan, Hilo, HI 96720, USA}\vspace{-0.5em}\\
\small{$^{2}$University of Texas-San Antonio, San Antonio, TX 78006, USA}\vspace{-0.5em}\\
\small{$^{3}$Eureka Scientific, Oakland, CA 94602, USA}\vspace{-0.5em}\\
\small{$^{4}$Department of Physics, University of California, Santa Barbara, Santa Barbara, California 93106, USA}\vspace{-0.5em}\\
\small{$^{5}$Department of Astrophysical Sciences, Princeton University, Princeton, NJ 08544, USA}\vspace{-0.5em}\\
\small{$^{6}$Department of Astronomy, University of Texas-Austin, Austin, TX 78712, USA}\vspace{-0.5em}\\
\small{$^{7}$Astrobiology Center, Osawa, Mitaka, Tokyo 181-8588, Japan}\vspace{-0.5em}\\
\small{$^{8}$Steward Observatory, The University of Arizona, Tucson, AZ 85721, USA}\vspace{-0.5em}\\
\small{$^{9}$National Astronomical Observatory of Japan, Osawa, Mitaka, Tokyo 181-8588, Japan}\vspace{-0.5em}\\
\small{$^{10}$Department of Astronomy, Graduate School of Science, The University of Tokyo, Tokyo 113-0033, Japan}\vspace{-0.5em}\\
\small{$^{11}$Department of Astronomy, Columbia University, New York, NY, USA}\vspace{-0.5em}\\
\small{$^{12}$Department of Physics and Astronomy, State University of New York-Stony Brook, Stony Brook, NY 11790, USA}\vspace{-0.5em}\\
\small{$^{13}$Department of Physics, University of Notre Dame, Notre Dame, IN 46556, USA}\vspace{-0.5em}\\
\small{$^{14}$NASA-Goddard Space Flight Center, Greenbelt, MD 20771, USA}\vspace{-0.5em}\\
\small{$^{15}$National Research Council-Herzberg, Victoria, BC V9E 2E7, Canada}\vspace{-0.5em}\\
\small{$^{16}$Department of Physics and Astronomy, University of Victoria, Victoria, BC V8W 2Y2, Canada}\vspace{-0.5em}\\
\small{$^{17}$Sydney Institute for Astronomy, School of Physics, University of Sydney, Australia}\vspace{-0.5em} \\
\small{$^{18}$Infrared Processing and Analysis Center, California Institute of Technology, Pasadena, CA 91125, USA}\vspace{-0.5em}\\
\small{$^{19}$Department of Astronomy, California Institute of Technology, Pasadena, CA 91125, USA}\vspace{-0.5em}\\
\small{$^{20}$Universite Cote d'Azur, Observatoire de la Cote d'Azur, Laboratoire Lagrange, Nice 06000, France}\vspace{-0.5em}\\
\small{$^{21}$Department of Mechanical Engineering, Princeton University, Princeton, NJ 08544, USA}\vspace{-0.5em}\\
\small{$^{22}$Department of Astronomy, Stockholm University, Stockholm 114 19, Sweden}\vspace{-0.5em}\\
\small{$^{23}$Department of Physics and Astronomy, George Mason University, Fairfax, VA 22030, USA}\vspace{-0.5em}\\
\small{$^{24}$Institute for Astronomy, University of Hawai`i, Hilo, HI 96720, USA}\vspace{-0.5em}\\
\small{$^{25}$Nicolaus Copernicus Astronomical Center, Polish Academy of Sciences, Torun 87-100, Poland}\vspace{-0.5em}\\
}
\date{}

\begin{document} 
\maketitle 
\begin{sciabstract}
 Direct imaging of gas giant exoplanets provides key information on 
 planetary atmospheres and the architectures of planetary systems.  However, few planets have been detected in blind surveys used to achieve imaging detections. 
Using Gaia and Hipparcos astrometry we identified dynamical evidence for a gas giant planet around the nearby star HIP 99770 and then confirmed this planet by direct imaging with the Subaru Coronagraphic Extreme Adaptive Optics Project.  HIP 99770 b orbits 17 astronomical units from its host star, with an insolation comparable to Jupiter's and a dynamical mass of 13.9--16.1 Jupiter masses. Its planet-to-star mass ratio (7--8$\times$10$^{-3}$) is comparable to that other directly-imaged planets.   The planet’s atmosphere resembles an older, less-cloudy analogue of the atmospheres of previously-imaged
exoplanets around HR 8799. 
\end{sciabstract}




Adaptive optics-assisted ground based telescopes have provided direct imaging detections of about 20 extrasolar gas giant planets 
\cite{Marois2010,Lagrange2010,Carson2013,Currie2014,Macintosh2015,Chauvin2017,Keppler2018,Currie2022}, 
These detections draw from so-called \textit{blind} (i.e. unbiased) surveys, where targets are selected based on system properties like age and distance.   However, the low yields of these blind surveys have shown exoplanets detectable using current direct imaging instruments are rare \cite{Nielsen2019}.  

Direct imaging provides constraints on an exoplanet's atmospheric properties like temperature, surface gravity, clouds, and composition \cite{Konopacky2013}.
But direct imaging data by themselves do not directly measure a planet's mass.  Masses reported for directly-imaged planets are inferred through luminosity evolution models, but these models have poor observational constraints and rely on often-uncertain host star ages \cite{Spiegel2012}.  
The typically wide separations and short temporal coverage for the locations of imaged exoplanets (i.e. their astrometry) can also lead to poor constraints on orbital parameters derived purely from direct imaging data alone \cite{Bowler2020}.   The dearth of direct imaging detections and poor mass and orbital constraints for imaged companions impede our understanding of gas giant (i.e. jovian) exoplanet atmospheres and the architectures of planetary systems.


Using an indirect detection method to guide direct imaging searches could (in principle) improve discovery yields and better constrain the atmospheres, orbits, and masses for a large population of exoplanets.  The few previous surveys that an indirect detection method to guide direct imaging observations used long-term radial-velocity (RV) trends to select direct imaging targets.  They imaged stellar companions and intermediate/high-mass brown dwarfs but not planets \cite{Crepp2016} (see Supplementary Text).  Alternatively, monitoring of a stars proper motion across the sky -- i.e. its astrometry -- can identify which stars are undergoing a proper motion acceleration caused by an unseen planetary-mass companion.   Furthermore, astrometric data can identify evidence for planetary-mass companions around young stars, which are unsuitable for precise RV measurements but are the best targets for imaging self-luminous gas giant planets \cite{Brandt2019}.  Combining the imaged planet's relative astrometry -- i.e. its location with respect to the host star -- with the host star's absolute astrometry can yield precise, directly-determined planet masses and improved constraints on orbital properties \cite{Brandt2019}.  The micro-arcsecond precision of the European Space Agency's Gaia mission combined with measurements 25 years prior from Hipparcos is sufficient to enable the astrometric detection of superjovian planets at Jupiter-to-Neptune like separations around the nearest stars.

\section*{Astrometry of HIP 99770}
HIP 99770 (also categorized as HD 192640 and 29 Cygni) is a chemically-peculiar star with an A spectral type, a distance of $d$ $\sim$ 40.74 $pc$, an effective temperature of $\approx$8000 K, luminosity of $\approx$13.9 $L_{\rm \odot}$, and mass of $\approx$1.7--2.0 $M_{\rm \odot}$\ \cite{supplements}.   
Different analysis methods -- e.g. kinematics, Hertzsprung-Russell diagram positions, and astroseismology data analyzed from the Transiting Exoplanet Space Satellite (TESS) -- give ages of either $\sim$40 Myr or 115--414 Myr old.
Data from the WISE, Spitzer Space Telescope, and Herschel Space Observatory show that the star is surrounded by a luminous, cold debris disk detected at far-infrared (IR) wavelengths extending to $>$ 150 astronomical units (au) \cite{supplements}.

As a part of a joint astrometry and direct imaging search for young exoplanets, we investigated evidence that HIP 99770 shows an astrometric acceleration due to an unseen companion: a deviation from linear motion across the sky.  We used astrometry from the Hipparcos-Gaia Catalogue of Accelerations (HGCA) \cite{Brandt2021}, a cross-calibration of the Hipparcos and Gaia missions.  HIP 99770's average proper motion between the Hipparcos and Gaia missions differs from the proper motion measured around 2016 in both the early (Gaia Data Release 2, hereafter Gaia DR2) and most recent and more precise (Gaia Early Data Release 3, hereafter Gaia eDR3) reduction \cite{GaiaeDR32021}.  The HGCA utilizing Gaia EDR3 measurements yields $\chi^{2}$ $\sim$ 7.23 for constant linear motion, revealing a statistically significant acceleration ($\approx$2.21 Gaussian sigma) at $>$97.31\% confidence or a false positive rate of $<$2.69\%.  The earlier DR2-version of HGCA reveals a comparably-significant acceleration ($\chi^{2}$ $\sim$ 9.80).  Other catalogs adopting independent analyses of Hipparcos and Gaia-eDR3 data also list HIP 99770 as having a statistically significant astrometric acceleration \cite{Kervella2022}. The astrometric acceleration due to this secondary companion ($sec$) depends on its mass $M$ and projected separation $R$, yielding a lower limit to $M_{\rm sec}/R_{\rm sec}^2$.  This ratio is consistent with a $\approx$11\,$M_{\rm Jup}$ object at a fiducial separation of $0.\!\!''5$ ($\approx$20\,au).  Previous high-contrast imaging observations with the Gemini Observatory rule out a stellar or substellar companion at wide separations ($\rho > 1''$) \cite{supplements}.

\section*{Direct imaging observations of HIP 99770}
Motivated by evidence for an astrometric acceleration, we observed HIP 99770 with the Subaru Coronagraphic Extreme Adaptive Optics Project (SCExAO) coupled to the Coronagraphic High-Resolution Imager and Spectrograph (CHARIS)\cite{Jovanovic2015,Groff2016}, in low-resolution (broadband) mode in 22 spectral channels covering the major near-infrared (IR) passbands simultaneously ($\lambda$ = 1.16--2.37 $\mu$m).  Our first two data sets consisted of two shallow observing sequences in July and September 2020.   
 Between May and October 2021, we conducted three deeper follow-up SCExAO/CHARIS observations and obtained one complementary data set in the thermal IR -- the $L_{\rm p}$ filter centered on $\lambda$ = 3\farcs{}78 $\mu$m -- with the NIRC2 camera on the 10-meter Keck II telescope.   
 
 SCExAO/CHARIS images show a faint point source, hereafter HIP~99770~b, located $\rho$ $\sim$ $0.\!\!''43$--$0.\!\!''44$ southeast of its host star (Figure 1; Table S1).  
 In the highest-quality data, HIP 99770 b is visible across the entire spectral range using advanced point-spread function (PSF) subtraction methods to remove light from the star combined with different observing strategies.  Each CHARIS data set shows a clear detection, providing astrometry at epochs spanning more than one year.  We also recover HIP 99770 b in the Keck/NIRC2 thermal IR data.


We use the direct imaging astrometry to reject the possibility that HIP 99770 b is a stationary background object at the $>$15-$\sigma$ level.  Between our first epoch (29 July 2020) and fourth epoch (13 July 2021), a background star should appear to move by $\sim$65 milli-arcseconds (mas) to both the west and south (Table S1) due to the star's proper motion but HIP 99770 b moved 23 $\pm$ 6 mas to the east and 29 $\pm$ 6 mas to the north--- in the opposite direction expected for a background star.  To masquerade as an orbiting companion, a distant background object would require an even higher proper motion than HIP 99770 A.  We reject the possibility that HIP 99770 b is a non-zero proper motion background star at the $>$5.7-$\sigma$ level from astrometry alone and at a $>$7.4-$\sigma$ level from astrometry and the planet's spectral features considered jointly \cite{supplements}.

\section*{Planet orbit and mass}

To determine HIP 99770 b's orbital properties and mass, we simultaneously fit its relative astrometry (from imaging data) and the primary's proper motions and anomalies (from the Gaia and Hipparcos astrometry \cite{Brandt2021}) using {\tt orvara}, a Markov Chain Monte Carlo (MCMC) code \cite{Brandt2021,ForemanMackey2013}. We used parallel tempering MCMC, in which progressively ``hotter" chains are more accepting of poor fits to the data to enable them to effectively explore multimodal posterior distributions \cite{Vousden2016}.  The coldest chain, i.e. the one least accepting of poor fits, is suitable for statistical inference.  We used 20 temperatures, each with 100 walkers taking 150,000 steps, to map the joint posterior of the orbital parameters.  We saved every 50th step and discarded the first half of the coldest chain as burn-in. 

For our baseline simulation, we conservatively adopted a uniform prior on the companion mass, $M_{\rm p}$, since this prior tends to yield higher companion masses than the standard log-normal prior (1/$M_{\rm p}$), and a Gaussian $1.8 \pm 0.2 M_{\rm \odot}$ prior on the primary.  For all other parameters, we adopted the standard uninformative priors \cite{Brandt2021d}.  Adopting a 1/$M_{\rm p}$ companion prior or changing the prior for the primary mass resulted in a lower companion mass but otherwise had little effect on our results \cite{supplements}.


The posterior probability distributions produced by {\tt orvara} for HIP 99770 b constraints orbital parameters 
(Figure 2).  For our baseline simulation, we derive a semimajor axis of 16.9$^{+3.4}_{-1.9}$ au, similiar to HR 8799 e's orbit \cite{Marois2010}.  HIP~99770~b's orbital eccentricity ($e$ = 0.25$^{+0.14}_{-0.16}$) is more consistent with values for directly imaged planets than brown dwarfs \cite{Bowler2020}.  We also estimate a mass for the primary of 1.85 $\pm$ 0.1 $M_{\odot}$.

HIP~99770~b has a dynamical mass of 16.1$^{+5.4}_{-5.0}$ Jupiter masses ($M_{\rm Jup}$) in our baseline simulation, yielding a planet-to-star mass ratio of $q = (8.4^{+2.8}_{-2.6}) \times 10^{-3}$.  
Adopting a 1/$M_{\rm p}$ prior on the companion mass instead of a uniform prior yields a lower mass of 13.9$^{+6.1}_{-5.1}$ $M_{\rm Jup}$ (so $q = (7.3^{+3.2}_{-2.7}) \times 10^{-3}$).  HIP 99770 b's mass is lower than the empirical mass separation between massive planets and low-mass brown dwarfs \cite{supplements}.
HIP 99770 b's mass ratio is similar to those for some imaged exoplanets like HR 8799 e ($q$ $\approx$ 6$\times$10$^{-3}$) and TYC 8998-760-1 b ($q$ $\approx$ 0.01--0.015) but is much lower than mass ratios measured for the brown dwarfs GJ 758 B and HD 33632 Ab ($q$ $>$ 0.04) \cite{Brandt2021c,Brandt2021d,Bohn2020}.

\section*{Planet atmospheric properties}
We compare HIP 99770 b's CHARIS spectrum (Data S1) to spectral templates \cite{Cruz2018} and the Montreal Spectral Library \cite{Gagne2014} 
to estimate its temperature.  HIP 99770 b lies at the transition from cloudy, methane poor L-type substellar objects to (nearly) cloud-free, T-type objects showing methane absorption (known as the ``L/T transition'').  The L7 template best reproduces its spectrum amongst all spectral templates.   In the spectral library, the L9.5 field dwarf SIMP J0956081-144706 (SIMPJ0956-1447) best reproduces HIP 99770 b's spectrum.
Thus, we assign HIP 99770 b a spectral type of L7--L9.5, which corresponding to an effective temperature of $T_{\rm eff}$ $\sim$ 1300--1500 K \cite{Stephens2009}.  

HIP 99770 b's atmosphere appears intermediate between the L9.5 field dwarf SIMPJ0956-1447 and the directly imaged, extremely cloudy L/T transition exoplanet HR 8799 d (Figure 3, top panel).  Thick clouds in the atmospheres of L/T transition objects result in a planet photosphere -- where the optical depth, $\tau$, is $\approx$1 --  more uniform with wavelength, resulting in a spectrum that is flatter, more blackbody-like and redder from $J$ band (1.25 $\mu m$) to $H$ (1.65 $\mu m$) to $K$ band (2.16 $\mu m$) than for an object with thinner clouds \cite{Currie2011}.  HIP 99770 b's spectrum is flatter and slightly redder than SIMPJ0956-1447.  However, it is more peaked than HR 8799 d's spectrum at $J$ and $H$ by a factors of $\sim$2 and $\sim$1.5 and has a bluer $J$--$K$ color, suggesting a cloud thickness intermediate between that of SIMPJ0956-1447's and HR 8799 d's.  At 2.2--2.4 $\mu m$, HR 8799 d lacks $CO$ absorption due to disequilibrium chemistry \cite{Konopacky2013}.  As HIP 99770 b's spectrum is fainter, the planet's atmosphere is likely closer to being in chemical equilibrium.

We compared the spectrum and photometry of HIP 99770 b to two grids of atmospheric models spanning a range of temperatures and surface gravities: the Lacy/Burrows grid (Data S2) and the BT-Settl grid \cite{supplements,Currie2011,Allard2012}.  These models adopting different prescriptions for clouds and atmospheric dust.   Atmospheric models require the presence of clouds to match HIP 99770 b's spectrum.  The best-fitting models from the Lacy/Burrows grid are less cloudy and less dusty than those that were successful in fitting the HR 8799 planets \cite{Currie2011}.
The best-fitting models have a surface gravity of log($g$) = 4--4.5 and temperatures of 1300-1600 $K$, where models with $T_{\rm eff}$ = 1300--1400 $K$ and log(g) = 4--4.5 provided the best fits (Figure 3B; \cite{supplements}).  Considering all models, the 2-$\sigma$ confidence intervals span log(g) = 4 to 5 and temperatures of 1250 to 1600 $K$.  
HIP 99770 b's best-fitting luminosity is log($L$/$L_{\rm \odot}$) = -4.53 $\pm$ 0.02, corresponding to a luminosity ratio of 2$\times$10$^{-6}$ relative to the primary.   Masses inferred from luminosity evolution models are consistent with the planet's dynamical mass if the planet is $\sim$80--200 Myr old \cite{supplements}.

\section*{Comparisons to other systems}
We compare the architecture of the HIP 99770 system to the Solar System.   HIP 99770 b orbits at $\sim$16.9 au, intermediate between the distances of Saturn and Uranus from the Sun.  The system's cold debris disk is likely $>$150 au from the star, about 3.5 times the typical distance from the Sun to Kuiper belt objects.   However, the 1.85 $M_{\odot}$ HIP 99770 is more luminous than the Sun.  Therefore, the amount of light that HIP 99770 b receives (i.e. its insolation) is similar to that received at $\sim$4.5 au in the Solar System, just interior to the orbit of Jupiter.   Likewise, HIP 99770's debris disk, if at 150 au, lies at a luminosity-scaled distance similar to that of the Kuiper belt from the Sun.   
Thus, like HR 8799 \cite{Marois2010}, the outer regions of the HIP 99770 system bear some characteristics of a scaled-up version of our own outer Solar System, albeit one where a single massive planet dominates.

The mass ratios ($q$) and orbital separations ($a_{\rm p}$) of substellar companions provide a coarse diagnostic of formation processes: the companion mass function for substellar objects reaches a local minimum at q $\sim$ 0.025, with smaller $q$ values mostly being planets formed in a disk, while larger values mostly being brown dwarf companions formed by molecular cloud fragmentation) \cite{Reggiani2016}.  
As with a few other RV-detected companions orbiting stars more massive than the Sun with masses at or slightly above the deuterium-burning limit of $\approx$ 13--14 $M_{\rm Jup}$, HIP 99770 b's mass ratio and separation ($q$ $\sim$ 0.0084, $a_{\rm p}$ $\sim$ 16.9 au) are more consistent with planets below the deuterium-burning limit detected by both direct imaging and RV (Figure 4).   A criteria based on the deuterium-burning limit itself fails to distinguish between planets and brown dwarfs: HIP 99770 b's mass is more consistent with planets \cite{supplements} (see Supplementary Text).

HIP 99770 b is an extrasolar planet jointly detection and characterized through direct and indirect techniques: direct imaging and precision astrometry.   
HIP 99770 b joins $\beta$ Pic b and HR 8799 e as imaged planets with both spectra and well-constrained dynamical masses (see Supplementary Text).   



\bibliography{bibliography}

\bibliographystyle{Science}

\section*{Acknowledgments}
We thank the three anonymous referees for their helpful comments and suggestions, which improved the quality of this paper.  We thank the Subaru Time Allocation committee, NASA-Keck Time Allocation Committee, and recent Subaru Director Michitoshi Yoshida for their support of this program through Open Use and Director's Discretionary Time allocations.  This research is based on data collected at Subaru Telescope, which is operated by the National Astronomical Observatory of Japan.  The authors acknowledge the very significant cultural role and reverence that the summit of Maunakea holds within the Hawaiian community.  We are most fortunate to have the opportunity to conduct observations from this mountain.   We thank Jonathan Gagne, Eric Mamajek, Richard Gray, and Jeremy Jones for extensive and helpful conversations regarding HIP 99770's properties.  This research has made use of the NASA Exoplanet Archive, which is operated by the
California Institute of Technology, under contract with the National Aeronautics and Space
Administration under the Exoplanet Exploration Program.  The development of SCExAO was supported by the Japan Society for the Promotion of Science (Grant-in-Aid for Research \#23340051, \#26220704, \#23103002, \#19H00703 and \#19H00695), the Astrobiology Center of the National Institutes of Natural Sciences, Japan, the Mt Cuba Foundation and the director’s contingency fund at Subaru Telescope.  \\
\textbf{Funding} TC was funded under NASA/XRP program NNX17AF88G and NASA-Keck Principal Investigator Award (Program 2020A$\_$N027). BL is supported by a 51 Pegasi Postdoctoral Fellowship.  CM acknowledges the support of the Natural Sciences and Engineering Research Council of Canada (NSERC).  SJM was supported by the Australian Research Council (ARC) through Future Fellowship FT210100485.  VD acknowledges support from NASA grant \#80NSSC19K0336 and Heising-Simons Foundation grant number \#2020-1823.
MT is supported by JSPS KAKENHI grant \#18H05442.\\
\textbf{Author Contributions} TC proposed and conducted the observations, reduced data, and wrote most of the paper.  GMB and TDB assisted with target selection and performed the joint direct imaging and astrometry analysis.  BL and AB generated planet atmosphere models.   RYL and SS performed and ran separate data reductions; TT contributed to data reduction.  OG, JL, VD, SV, NS, JC, TG, and TK operated Subaru's facility AO system, SCExAO, and CHARIS during observations.  CM and WT reduced and analyzed archival Gemini data.   SJM performed astroseismology analysis.  MK, TU, MM, MJ, JW, KH, JN, KH, and JK contributed to proposals and draft comments.  NJ, FM, NJK, and MH developed, commissioned, and maintained SCExAO and CHARIS and contributed to paper comments.  MT oversaw group collaboration, proposals, and draft circulation.\\
\textbf{Competing interests}  The authors declare no competing interests.\\
\textbf{Data and materials availability}  Raw CHARIS data are available for public download from the Subaru SMOKA archive --  https://smoka.nao.ac.jp/ .  Keck data are available from the Keck Observatory Archive (https://koa.ipac.caltech.edu/cgi-bin/KOA/nph-KOAlogin).   WISE mission, Spitzer Space Telescope, and Herschel Space Observatory data are available from the NASA/IPAC Infrared Science Archive -- https://irsa.ipac.caltech.edu/applications/Gator/ -- where the Spitzer data used are from the Spitzer Enhanced Imaging Products data set.
Gemini data are available from the Gemini Observatory archive: https://archive.gemini.edu/searchform.  TESS data are available from the Mikulski Archive for Space Telescopes at the following location: https://mast.stsci.edu/portal/Mashup/Clients/Mast/Portal.html.  In all cases, data can be found by searching for the star's name (HIP 99770).  
The HIP 99770 b spectrum used for analysis and the Lacy/Burrows atmospheric models used to interpret this spectrum are available as Data S1 and Data S2, respectively, in the supplements.

 \newpage

  \begin{figure*}[h!]
    \includegraphics[width=1.0\textwidth]{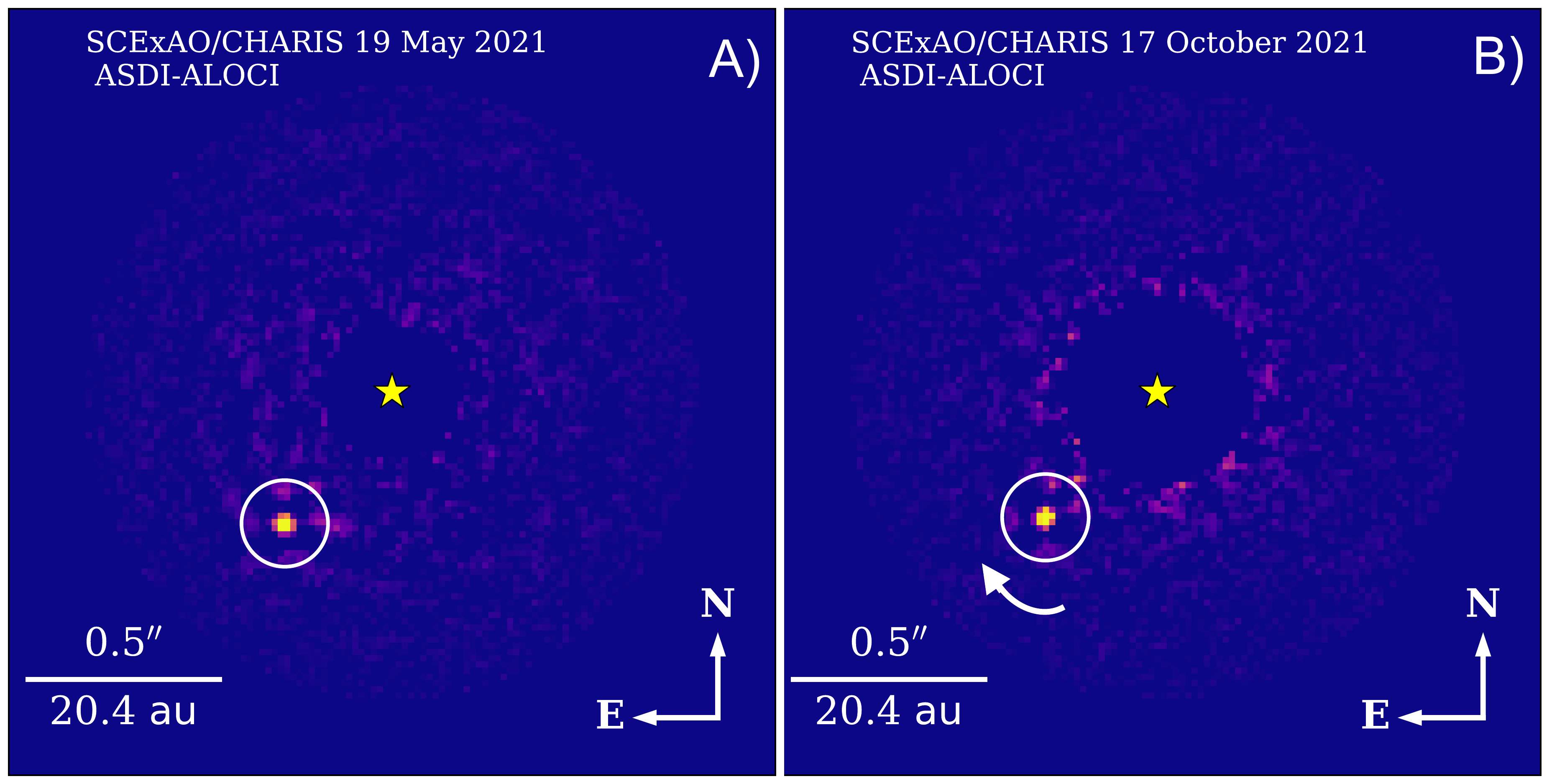} 
   \vspace{-0.325in}
  \caption{\textbf{Direct images of HIP 99770 b using SCExAO/CHARIS from (A) 19 May 2021 and (B) 17 October 2021}: our two highest-quality data sets.   We used the Adaptive, Locally-Optimized Combination of Images algorithm (ALOCI) in combination of angular differential imaging (ADI) and spectral differential imaging SDI (ASDI) to remove the stellar halo light \cite{Currie2012,Marois2006,SparksFord2002}.   HIP 99770 b is identified by a white circle.  The white arrow in the (B) panel shows the direction of HIP 99770 b's orbital motion.  The star symbol denotes the host star's location.   The color intensity scalings are linear with a minimum of zero and maximum scaled to the mean signal within a PSF core.}
    \vspace{-0.in}
\end{figure*}

 \begin{figure*}[h!]
    \includegraphics[width=1.0\textwidth]{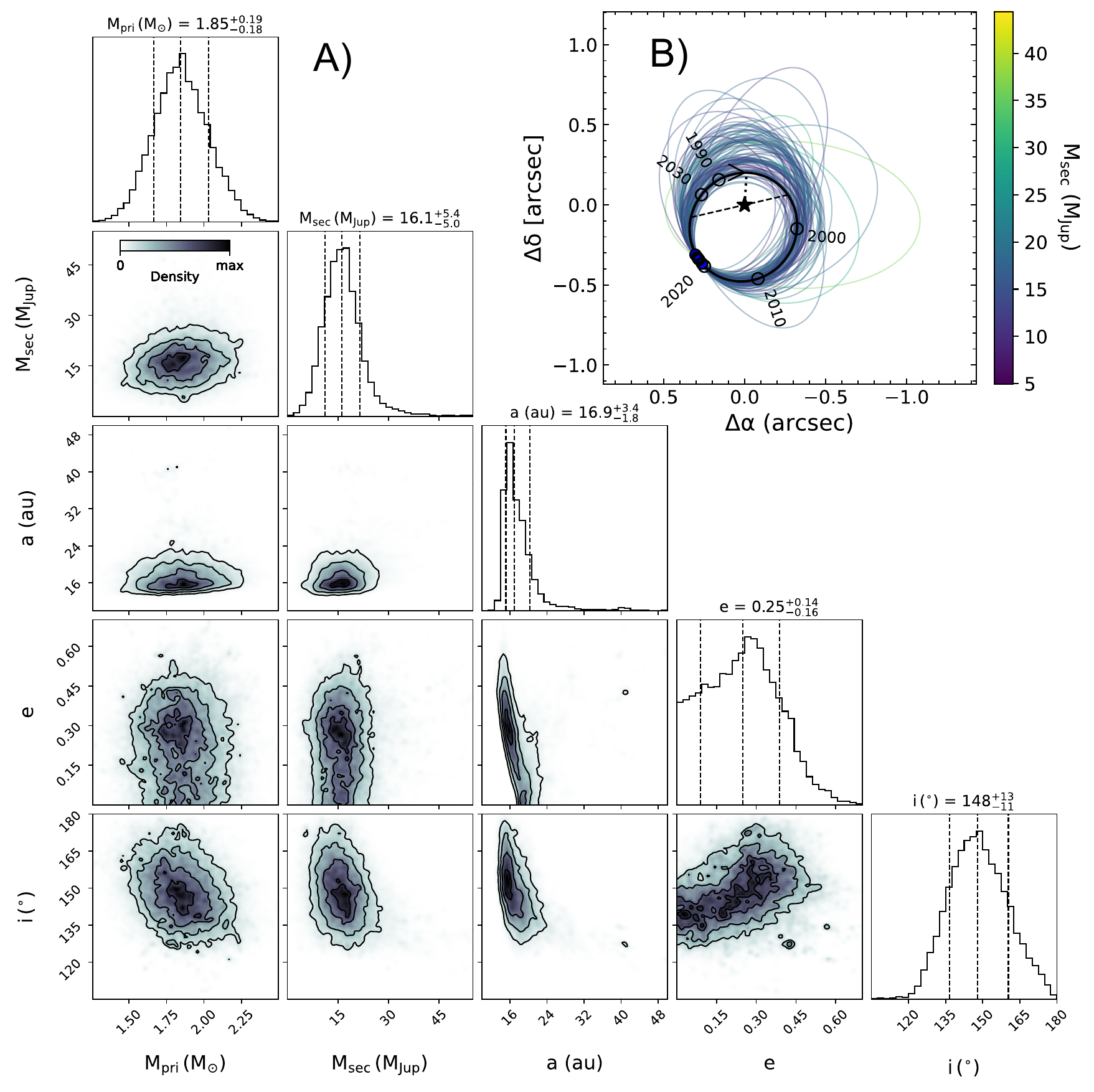} 
   \vspace{-0.2in}
  \caption{\textbf{Astrometric analysis for HIP 99770 b}.  The corner plot (A) shows posterior probability distributions of key orbital parameters, the mass of HIP 99770 b and the mass of the HIP 99770 primary.
    The (B) panel displays the best-fit orbit along with 100 orbits drawn from our posterior probability distribution, color-coded by HIP 99770 b's mass.   The black curve corresponds to the best-fitting orbital model.}
    \vspace{0.1in}
    \label{fig:orbits}
\end{figure*}

 \begin{figure*}[h!]
  \centering
  \includegraphics[width=0.95\textwidth,clip]{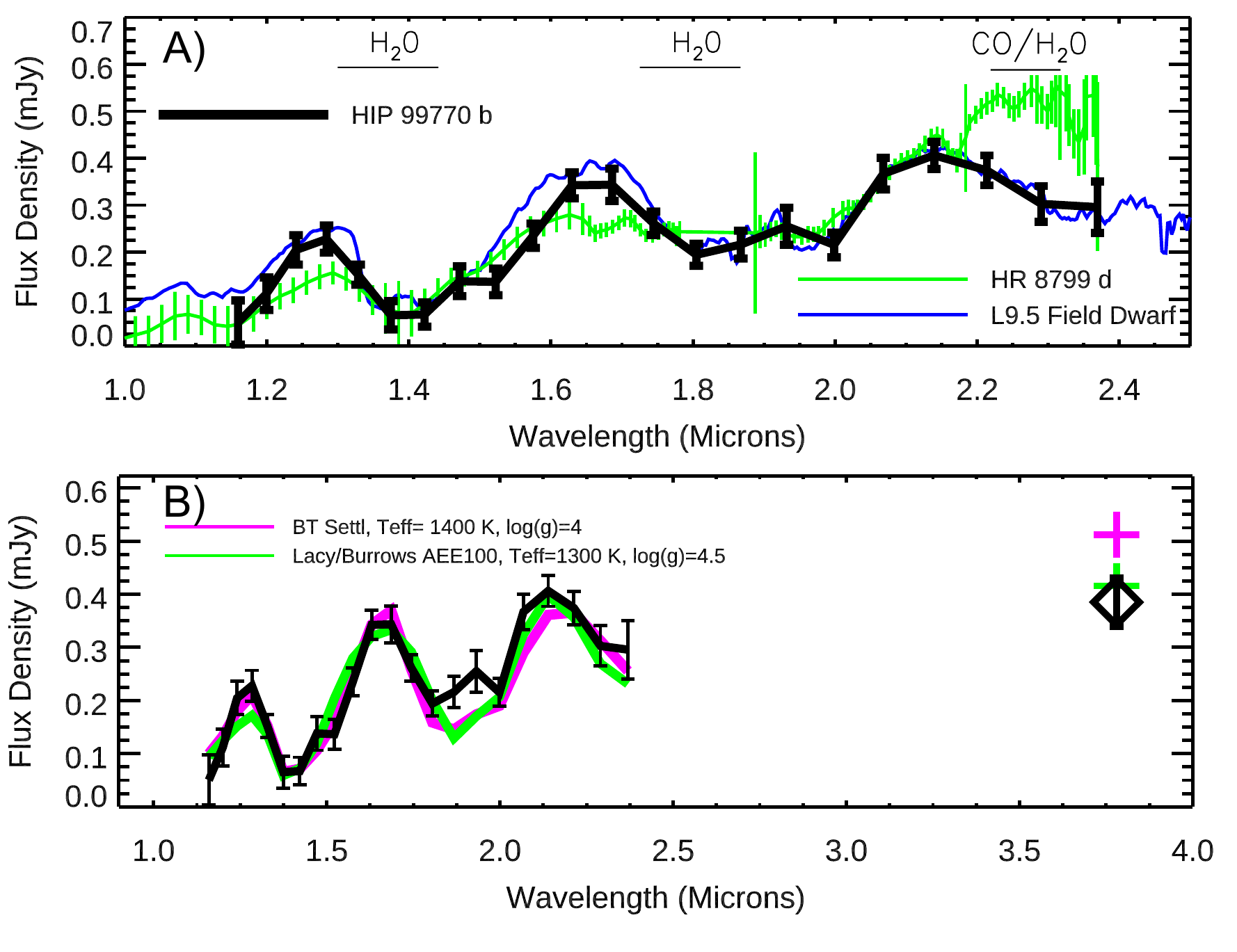} 
   \vspace{-0.2in}
  \caption{\textbf{Atmospheric characterization of HIP 99770 b.}  (A) HIP 99770 b spectrum compared to spectra for HR 8799 d \cite{Greenbaum2018,Zurlo2016} and the L9.5 field dwarf SIMPJ0956-1447 (drawn from the Montreal Spectral Library), normalized to HIP 99770 b's spectrum at 2.1 $\mu m$.   Horizontal bars show the wavelength ranges for absorption due to $H_{\rm 2}O$ and $CO$.  (B) HIP 99770 b's spectrum (black line) and photometry (black diamond) compared to representative, well-fitting atmospheric models from the BT-Settl (magenta) and Lacy/Burrows (green) grids ($\chi^{2}_{\rm \nu}$ = 1.345 and 1.397, respectively) \cite{Allard2012}.   Lines correspond to model predictions for spectra; plus signs correspond to predicted $L_{\rm p}$ photometry.  The best-fit radii and implied masses are 0.92 $R_{\rm J}$, 3.3 $M_{\rm Jup}$ and 1.05 $R_{\rm J}$, 13.5 $M_{\rm Jup}$, respectively.   Both models include clouds and atmospheric dust. The Lacy/Burrows models include non-equilibrium carbon chemistry and updated opacities.  The spectra of HIP 99770 b and other objects are in flux density units of milli-Janskys (mJy).   All error bars represent 1-$\sigma$ uncertainties.  The Lacy/Burrows model is labeled AEE100, which corresponds to a model atmosphere with intermediate cloudiness and a modal size of particles entrained in these clouds of 100 $\mu m$: see \cite{supplements,Burrows2006,Currie2011,Madhusudhan2011} for more details.   
  }
    \vspace{-0.2in}
    \label{fig:atmosmodels}
\end{figure*} 

 \begin{figure*}[h!]
  \centering
  \includegraphics[width=0.95\textwidth,clip]{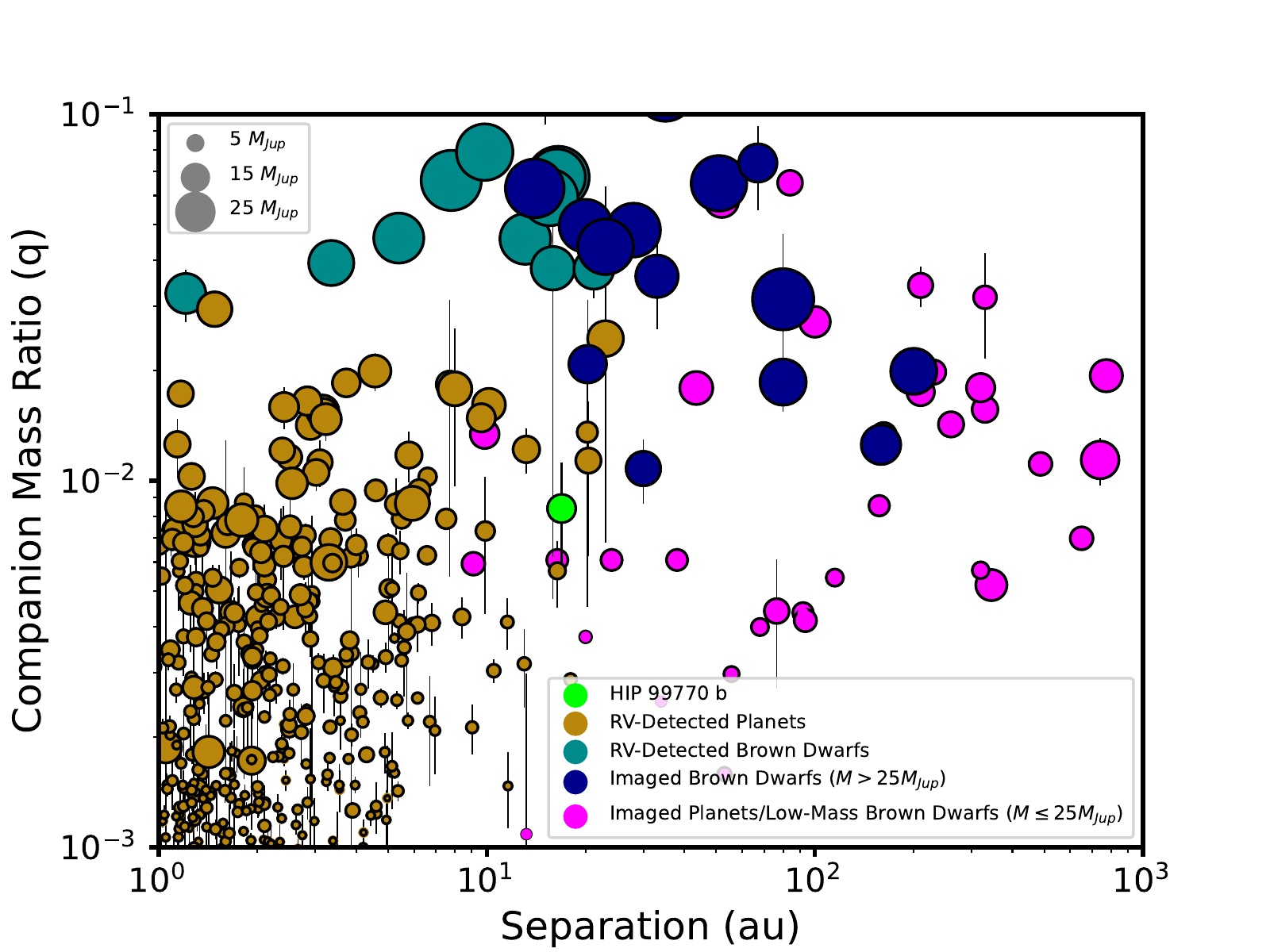} 
   \vspace{-0.1in}
  \caption{\textbf{Mass ratio vs. separation for substellar companions.}  Data are from the NASA Exoplanet Archive \cite{exoarchive} (https://exoplanetarchive.ipac.caltech.edu/) and exoplanet.eu\cite{Schneider2011} with small modifications (see \cite{supplements}).   We show only objects detected by the direct imaging and radial-velocity techniques.  Symbols sizes are proportional to the object mass.  HIP 99770 b lies well within the region contiguous with RV-detected and directly-imaged exoplanets.
  }
    \vspace{-0.2in}
    \label{fig:massratio}
\end{figure*}



\clearpage
\noindent{\textbf{\large{Supplementary Materials}:}}\\
Materials and Methods\\
Supplementary Text\\
Figures S1 to S14\\
Tables S1 to S6\\
Data S1\\
Data S2

\newpage

\section*{Materials and Methods}

\textbf{Systems Properties}: The research literature contains inconsistent entries for some of HIP 99770's properties (Table S1), especially its spectral type and temperature estimates.   Some studies classified HIP 99770 A as an A0V--A2V star based primarily on its weak metal lines \cite{vanBelle2009}.   Complicating its spectral classification, HIP 99770 is a $\lambda$ Boo star whose spectrum is depleted in metals \cite{Slettebak1952,Gray1988}.  Its Balmer sequence is more consistent with a mid A star, and the star's spectral type has also been listed as kA1.5hA7mA0.5 \cite{Murphy2017}.     The star's $\lambda$ Boo status explains its weak metal lines, though weak metal lines are not necessarily indicative of an overall low bulk metallicity.

 A star's relative photometry in the $V$ ($\lambda_{\rm c}$ = 0.55 $\mu m$) and $K_{\rm s}$ ($\lambda_{\rm c}$ = 2.16 $\mu m$) photometric passbands provides a coarse estimate of its spectral type \cite{Pecaut2013}.  Relative photometry between the blue ($B_{\rm P}$) and red ($R_{\rm P}$) passbands in Gaia can also be used to estimate spectral type \cite{GaiaeDR32021}.
HIP 99770 has a $V$-$K_{\rm s}$ color of 0.518 mag.  Its color in the Gaia passbands is $B_{\rm P}$-$R_{\rm P}$ $\sim$ 0.26 mag \cite{GaiaeDR32021}.   These values are redder than expected for an A0--A2 star.  Using $V$-$K_{\rm s}$ to estimate reddening \cite{Cardelli,Pecaut2013} implies $E(B-V)$ = 0.17, 0.12, 0.10 and 0.03 for A0, A2, A3, and A5 spectral types.   Values quoted for HIP 99770's E(B-V) vary in the literature, ranging from 0.014 to 0.025 \cite{Murphy2017}.   The star shows Na absorption consistent with gas absorption at a level of E(B-V) $\approx$ 0.02 \cite{Lallement2003}.   
Another analysis \cite{Murphy2017} found the star's effective temperature to be $T_{\rm eff}$ $\sim$ 7960 $K$ consistent with an A5--A6 star reddened by a small amount (E(B-V) $\sim$ 0.02) but inconsistent with temperatures characteristic of early A stars \cite{Pecaut2013}.   

We therefore adopt a temperature of $T_{\rm eff}$ = 8000 $K$ and a spectral type matching this temperature: A5--A6.   Using a $M_{\rm V}$ to luminosity relationship \cite{Pecaut2013} we infer an apparent luminosity of 13.86 $L_{\rm \odot}$ and a mass of 1.84 $M_{\rm \odot}$.   Adopting instead a luminosity from a mass-luminosity relationship and using the mass range of 1.7--2.0 $M_{\odot}$ \cite{Jones2016}, we get a consistent range of $L_{\rm \star}$ = 8.4--16 $L_{\rm \odot}$.  The star is unsaturated in 2MASS $K_{\rm s}$ band but likely saturated in the 2MASS J and H filters.
We estimated its $J$, $H$, and $L_{\rm p}$ photometry assuming the colors of an A5V star.

HIP 99770's age diagnostics lead to seemingly conflicting conclusions.   On one hand, its stellar kinematics are consistent with members of the $\approx$40 Myr-old Argus association \cite{Zuckerman2011,Zuckerman2019}.   
   Its three-dimensional space velocity matches the mean velocities for the Argus core.  However, the star's physical separation from the Argus core is $\sim$ 80 $pc$, although some members are more separated \cite{Zuckerman2019}.   A kinematics analysis \cite{Gagne2014} updated with Gaia astrometry but not including photometry yields a 99.7\% likelihood that HIP 99770 is an Argus member.  Thus, the evidence for HIP 99770's kinematic membership in Argus is strong but not decisive.

On the other hand, the star's HR diagram position suggests an age older than Argus members, between that of the Pleiades and Ursa Majoris (115 Myr and 414 Myr, respectively \cite{Gaia2018,Jones2016}) (Figure S1).  HIP 99770 has a rotation rate of $v$ sini $\sim$ 65 km s$^{-1}$ \cite{Royer2007}: if viewed at a high inclination angle -- i.e. closer to pole-on -- the star's bolometric luminosity can be far lower than its apparent luminosity as was found for $\kappa$ And \cite{Jones2015}, moving its HR diagram position away from Ursa Majoris and much closer to the Pleiades sequence or possibly even below.   

An independent stellar age can be calculated using asteroseismology. HIP\,99770 pulsates as a $\delta$\,Sct variable star, a class for which asteroseismic ages are measureable \cite{murphyetal2021a}. Regular patterns are observed in the pulsation frequencies of young $\delta$\,Sct stars \cite{beddingetal2020}, with a characteristic large spacing ($\Delta\nu$) that corresponds to the square root of the mean stellar density ($\bar{\rho}$), i.e. $\Delta\nu\sim\sqrt{\bar{\rho}}$. As the star ages, its mean density decreases, and so too does its $\Delta\nu$. 

For our asteroseismic analysis, we downloaded TESS light curves from the Mikulski Archive Space Telescopes using the lightkurve package \cite{lightkurvecollaboration2018}.   We used Pre-search Data Conditioning Simple Aperture Photometry (PDCSAP) taken at 2-min cadence to construct an echelle diagram \cite{beddingetal2020}.
We measure $\Delta\nu = 4.86\pm0.03$, which suggests a much lower density than a typical slowly rotating $\delta$\,Sct star near the zero-age main sequence (e.g. Pleiades $\delta$\,Sct stars have $\Delta\nu\sim6.90$, \cite{murphyetal2022a}). We consider that this low density probably arises from centrifugal deformation due to rapid rotation, which can reduce the mean density by 37\% for a star rotating at 2/3 of its break-up velocity \cite{murphyetal2022a}. 

To evaluate the implications for the age of HIP\,99770, we constructed rotating stellar models for a mass of 1.8\,M$_{\odot}$, a solar metallicity ($Z=0.0144$; \cite{Asplund2009}), and a range of uniform initial rotation rates, $v_{\rm in}$ using the Modules for Experiments in Stellar Astrophysics (MESA) code \cite{Jermyn2022}. These correspond to: an equatorial viewpoint ($v_{\rm in} = v\sin i = 65$\,km\,s$^{-1}$); a coplanar viewpoint ($i=32^{\circ}$, $v_{\rm in} = 122$\,km\,s$^{-1}$); the 8th decile of the rotational velocity distribution for mid-A stars \cite{Royer2007}, being $v_{\rm in} = 240$\,km\,s$^{-1}$ and implying $i=16^{\circ}$; and a model at 280\,km\,s$^{-1}$, whose rotation is so rapid as to delay the pre-main sequence contraction -- this model eventually reaches 99.6\% of its break-up velocity at age 100\,Myr. The results are shown in Figure S2. The asteroseismological data are inconsistent with an equatorial viewpoint for HIP\,99770 and also with a coplanar viewpoint, both of which imply a density higher than that observed. Conversely, the inferred density is too high for HIP\,99770 to be rotating at the limit of break-up. An age of $\sim$180\,Myr, slightly older than the upper limit of the Pleiades age, is found if the star is rotating uniformly at 240\,km\,s$^{-1}$. Our calculations suggest a rotation rate of 180\,km\,s$^{-1}$ would be consistent with a UMa age of $\sim$414\,Myr.   

Considering all lines of evidence, we adopt a bimodal distribution for possible ages: 40 $Myr$ (based on Argus membership) or 115--414 Myr (based on the star's HR diagram position and astroseismology).  Despite HIP 99770's kinematics suggesting Argus membership, we favor the latter, older age range given the difficulty in reconciling the star's internal density with younger ages.   An age of 115--200 Myr provides a self-consistent picture given the star's properties and the age range consistent with HIP 99770 b's luminosity and dynamical mass (see below).  

To search for evidence of a debris disk around HIP 99770, we queried the NASA/IPAC Infrared Science Archive for photometric data from the Spitzer Space Telescope, WISE mission, and Herschel Space Observatory.   From this archive, HIP 99770 has photospheric colors at wavelengths shorter than $\lambda$ $<$ 22 $\mu m$, and thus no evidence for copious, warm circumstellar dust.  For example, the relative photometry between the WISE 3.4 $\mu m$ filter (W1) and the WISE 22 $\mu m$ filter (W4) is W1-W4 $\sim$ -0.016, which is consistent with a bare photosphere \cite{Pecaut2013}.  However, from 70 $\mu m$ data taken with the Photodetector Array Camera and Spectrometer instrument on Herschel (2012 April 11), HIP 99770 has a 70 $\mu m$ flux density of 157.5 $mJy$ $\pm$ 11.8 $mJy$.     If its 70 $\mu m$ emission originated only from a stellar photosphere, HIP 99770 should have had a flux density of $\approx$ 12 mJy.  Therefore, HIP 99770 has an observed to photospheric flux ratio of $\approx$ 13 at 70 $\mu m$, similar to stars with cold debris disks like AU Mic \cite{Rebull2008}.   Assuming a ring of blackbody dust peaking at $\lambda$ $>$ 70 $\mu m$, HIP 99770's debris disk is colder than $\sim$40 $K$ and at least 150 au (3\farcs{}7) in radius.  The Herschel images detect this disk: the flux attributed to it is not from a contaminating source (Figure S3).   


\textbf{High-Contrast Imaging Observations and Data Reduction}: We observed HIP 99770 as a part of our pilot accelerating star survey using Subaru/SCExAO \cite{Currie2021}.  Table S2 summarizes our HIP 99770 high-contrast imaging observations.
Typically, the seeing as measured in $V$ band was slightly poorer than the Maunakea median value ($\sim$ $0.\!\!''6$).  
The June 2021 Keck data were of the highest quality, with excellent seeing ($\sim$ $0.\!\!''3$--$\sim$ $0.\!\!''45)$, low winds, and extremely low precipitable water vapor levels ($<$ 1 mm).
Similarly, the May 2021 SCExAO observations typically had superb conditions, including extremely low winds, resulting in an long atmospheric coherence time: a circular dark hole was visible in each CHARIS wavelength slice, indicating deep raw contrasts for SCExAO.  However, the data periodically suffered from Low-Wind Effect \cite{Milli2018}, leaving us with 39 minutes worth of data, slightly less than half of the original amount.   Conditions for October 2021 were poorer than May/June 2021 but stable and better than other data sets.
We measured an $H$ band Strehl ratio of $\sim$0.80 at the start of the July 2021 sequence.   Raw contrasts were similar or slightly better for the October 2021 data, while they were $\sim$ 2--3 times deeper for the May 2021 data, similar to prior performances where the $H$ band Strehl ratio measured 0.90--0.95\cite{Currie2018a,Currie2020b}.   

To remove stellar halo light, we obtained all observations in angular differential imaging (ADI) mode, allowing the sky to rotate, while keeping the telescope pupil fixed on the detector \cite{Marois2006}.  The CHARIS integral field spectrograph data also enabled spectral differential imaging (SDI) \cite{Marois2000,SparksFord2002}.   For CHARIS, we utilized satellite spots produced by a 25 nm sine wave modulation on the SCExAO deformable mirror for astrometric and spectrophotometric calibration \cite{Jovanovic2015-astrogrids}.  To calibrate NIRC2 photometry, we took unsaturated images of HIP 99770.  The CHARIS and NIRC2 data utilized a Lyot coronagraph with an occulting mask of 0\farcs{}139 and 0\farcs{}3 in radius, respectively.

To extract CHARIS data cubes from raw data, we used the standard CHARIS pipeline \cite{Brandt2017}.  We utilized the CHARIS Data Processing Pipeline (DPP) \cite{Currie2020b} for subsequent reduction steps: i.e. sky subtraction, image registration, spectrophotometric calibration, spatial filtering, PSF subtraction, throughput correction, and spectral extraction.   
To process NIRC2 data, we used a general imaging pipeline \cite{Currie2011}, which performed the same reduction tasks.   

For PSF subtraction, we used the Adaptive, Locally Optimized Combination of Images (ALOCI) algorithm in combination with ADI or ADI+SDI (i.e. ASDI)  \cite{Currie2012,Currie2015}.   We explored a wide range of algorithm parameters, varying the \textit{singular value decomposition} (SVD) cutoff, the number of $N$ best-correlated frames included in the reference library, the rotation gap, the subtraction/optimization zone geometries, and the use/disuse of a pixel mask over the subtraction zone \cite{Lafreniere2007,Marois2010b,Currie2015}, etc.     

Figure S4 shows the resulting images in each epoch.   In the October 2021 and May 2021 data, HIP 99770 b is detected at a signal-to-noise ratio (SNR) of SNR = 39--45, where we calculate detection significances using standard metrics, correcting for small number statistics \cite{Currie2011,Mawet2014}.   For data processed using ADI only and in May/October 2021 for CHARIS and June 2021 NIRC2, HIP 99770 b is easily detected over the entire ALOCI parameter space explored, peaking at SNR = 24.9 for the October 2021 data.   As a separate comparison for select CHARIS data, we also detected HIP 99970 b using an implementation of the Karhunen-Lo\`eve Image Projection (KLIP) algorithm \cite{Soummer2012} with ADI: SNR = 17.8 for the October 2021 data.   For October 2021 data reduced using ALOCI-ADI or KLIP-ADI, the final, sequence-combined PSF subtracted cubes detect HIP 99770 b in every single CHARIS spectral channel.  Utilizing SDI yielded an additional factor of 1.8--2.3 gain in SNR in the final ALOCI reductions.

To estimate and correct for spectrophotometric and astrometric biasing due to processing, we followed standard forward-modeling methods \cite{Pueyo2016,Currie2018a}, comparing the input and output flux density in each channel and position in the wavelength collapsed images.   We assumed an L7 dwarf template spectrum inputed into each sequence, although the spectral type had no effect on the forward-model for ADI-reduced data.   
The astrometric offset was small, typically $\sim$ 0.05--0.25 pixels in both east and north.  We adopted astrometry from the ADI-only reductions because the astrometric biasing is independent of the input spectrum.   Our astrometric error budget considers the intrinsic SNR of the detection, a systematic error of 1/4 pixel in the centroid position, and an uncertainty in the pixel scale and north position angle of 0.05 mas and 0.27$^{o}$, respectively.   We assessed any evidence for a shift in the CHARIS north position angle offset from previously determined values \cite{Currie2018a} by obtaining multiple contemporaneous CHARIS observations and some NIRC2 observations of HD 1160 to detect its companion \cite{Nielsen2012}, but found no such evidence \cite{Kuzuhara2022}.   HIP 99770 b's May--July 2021 CHARIS and June 2021 NIRC2 astrometry are consistent within the uncertainties.  Table S3 records our astrometry and the SNR of our detection in each epoch.

We adopted the throughput-corrected spectrum from the October 2021 ADI-ALOCI reduction, because it was the highest SNR reduction without using SDI and yielded a planet detection in every CHARIS spectral channel (Figure S5).    Following previous work \cite{Greco2016}, we computed the spectral covariance from these data.   The covariance at HIP 99770 b's position is dominated by spectrally and spatially uncorrelated noise ($A_{\rm uncorr}$ $\sim$ 0.88).  Spectra extracted from the May 2021 ALOCI-ADI and October 2021 KLIP-ADI reductions are also consistent over most channels (Figure S6) but have lower SNRs and stronger off-diagonal terms in their covariance matrices leading to larger spectroscopic uncertainties.   Table S4 lists the measured and derived properties for HIP 99770 b: photometry in the $J$, $H$, $K_{\rm s}$, and $L_{\rm p}$ passbands, atmospheric properties, mass, mass ratio, and orbital properties. 

\textbf{Archival Imaging Data}: We searched for additional HIP 99770 data that could contain a prediscovery detection or otherwise constrain properties of HIP 99770 b.  
HIP 99770 was targeted as a part of the International Deep Planet Survey (IDPS), observed in the $CH_{\rm 4}$ filter ($\lambda$ = 1.63 $\mu m$) in 2008 with the NIRI camera on Gemini North telescope \cite{Galicher2016},   
though contrast curves and a list of point sources detected were not published.  If a background object, HIP 99770 b would appear at an East, North position of [E,N]\farcs{} $\approx$ [-0.6,-1.23] at a contrast of $\approx$ 1.5$\times$10$^{-5}$, which should be detectable from these data.  We reduced the NIRI observation using a Julia language pipeline \cite{Thompson2021}.    The NIRI data are saturated out to 0\farcs{}5: thus we have no sensitivity to planets at separations less than 0\farcs{}5.   While this pipeline achieves a contrast sufficient to recover HIP 99770 b if it were a background object at $\rho$ $>$ 0\farcs{}75 (1.2$\times$10$^{-5}$), we fail to see it in these data.  Although we did not consider the NIRI data in our orbit modeling, the NIRI non-detection of HIP 99770 b would also rule out some high eccentricity, high-mass orbital solutions.

\newpage
\section*{Supplementary Text}
\textbf{Testing Whether HIP 99770 b Can Be A Background Object}: 
We consider whether the HIP 99770 system's astrometry and HIP 99770 b' spectrum are compatible with a companion being a background object with a non-zero proper motion.

First, HIP 99770 b exhibits orbital motion roughly along the direction of the star's proper motion vector: a background object mimicking HIP 99770 b's astrometry must then have an even higher proper motion than the HIP 99770 primary.   To estimate the proper motion of a background star required to mimic this orbital motion, we follow a simplified version of the analyses performed for the background star formerly called HD 131399 Ab \cite{Nielsen2017}, determining the combination of background proper motions and parallaxes that best match HIP 99770 b using a $\chi^{2}$ analysis and estimating 1-$\sigma$ uncertainties on these values.  

We find that a star masquerading as HIP 99770 b must have a very high proper motion with a large parallax.   The best-fitting background star proper motion is $\mu_{\rm \alpha~\star}$, $\mu_{\rm \delta}$ $\sim$ 93, 100 $\pm$ 8, 9 mas yr$^{-1}$ with a parallax of $\pi$ $\sim$ 24 mas.  No solution with a parallax smaller than $\pi$ $\sim$ 9 mas can match HIP 99770 b's proper motion within the 5-$\sigma$ level.

We then use a Besancon population synthesis model \cite{Robin2003} to compare the expected distributions of background star proper motions and parallaxes to those required for HIP 99770 b to be a misidentified high proper motion background star.   We synthesize the background star distribution over a square-degree area with distances ranging from 0 kpc to 1 kpc ($\pi$ $>$ 1 mas).  Our population includes stars with $H$ band magnitudes as faint as $H$ = 19.47, i.e. an object with a contrast relative to HIP 99770 of 10$^{-6}$: slightly deeper than our 5-$\sigma$ contrast limit at HIP 99770 b's angular separation ($\sim$1.6$\times$10$^{-6}$).   The simulation synthesizes a total of 10,732 stars.

Even if parallax constraints are ignored, this synthesized population does not generate any background star whose kinematics $\nu$ match HIP 99770 b's at the 5-$\sigma$ level (Figure S7) \cite{Nielsen2017}).   Thus, the likelihood of stars matching HIP 99770 b's proper motion at the companion's angular separation of $\rho$ is $P(\nu|\rho)$ $<$1/10732 or $\le$9.3$\times$10$^{-5}$.   Only 0.1\% of stars in our simulation have parallaxes greater than 9 mas, although most higher proper motion stars should also have larger parallaxes.   We then compute \cite{Nielsen2017} the probability of a background star $BG$ also matching HIP 99770 b's 2-$\sigma$ range in angular separation, $\rho$: $P(BG)P(\rho|BG$).  For a 2-$\sigma$ range over our entire set of observations of $\rho$ $\sim$ 0\farcs{}43--0\farcs{}46 and assuming we would detect HIP 99770 b between 0\farcs{}25 and 1\farcs{}1 from the star, we find $P(BG)P(\rho|BG)$ $\sim$ 0.0030$\times$0.02326 $\sim$ 6.9$\times$10$^{-5}$.  Combining these two probabilities, the probability that HIP 99770 b is a background object ($P(BG)P(\rho|BG)P(\nu|\rho)$) is $<$6.5$\times$10$^{-9}$ ($>$5.7-$\sigma$).  For comparison, in the case of HD 131399 Ab, $\sim$ 0.89\% of the simulated population of background stars could match its proper motion and parallax: the estimated probability of HD 131399 Ab being a background star is $\sim$2.8$\times$10$^{-7}$ \cite{Nielsen2017}.   

We also compute the probability of a bound companion like HIP 99770 b and compare it to the background object probability to quantify whether HIP 99770 b is more likely to be a background star or a bound companion \cite{Nielsen2017}.  Assuming a frequency of 0.8\% for companions with masses between 13 $M_{\rm J}$ and 80 $M_{\rm J}$ \cite{Nielsen2019} and adopting a mass and semimajor axis distributions for the number $N_{\rm comp}$ of such companions of $dN_{\rm comp}/dM$ $\propto$ $M^{-0.47}$ and $dN_{\rm comp}/da$ $\propto$ $a^{-0.65}$, we estimate the probability of a 11--21.5 $M_{\rm J}$ planetary companion ($pl$) at HIP 99770 b's angular separation $\rho$ (0\farcs{}43--0\farcs{}46) of $P(pl)\times~P(\rho|pl)$ = 0.00192$\times$0.032 = 6.4$\times$10$^{-5}$.   Unlike the case of HD 131399 Ab, all of our orbital solutions are consistent with a gravitationally-bound object: the fraction of those with velocities consistent with HIP 99770 b's measurements within 2-$\sigma$ is therefore 0.95.  

The relative probability of HIP 99770 b being a background star instead of a bound companion is lower than $\frac{P_{\rm BG}}{P_{\rm pl}} \sim \frac{6.5 \times 10^{-9}}{6.1 \times 10^{-5}}$ $\sim$ 10$^{-4}$ ($\approx$ 3.7 $\sigma$).  In other words, based on kinematics alone, HIP 99770 b is 10,000 times more likely to be a planet than a background star.   In contrast, HD 131399 A"b" was found to be 43,000 times more likely to be a background star: HIP 99770 b's relative likelihood of being a planet is nearly half a billion times higher. 

Second, HIP 99770 b's spectrum is substellar with a L7--L9.5 spectral type.   We regard this evidence as far stronger than for HD 131399 Ab \cite{Wagner2016} because our SNR is higher and our spectrum covers the three major near-infrared passbands ($J$, $H$, and $K$: 1.1--2.4 $\mu m$)  instead of just the blue half of $H$ band, showing a sawtooth-like spectrum characteristic of a late L dwarf.   Following \cite{Macintosh2015} we compute the likelihood of such a late L dwarf background object, adopting the space density of L5--T0 dwarfs \cite{Reyle2010} of 2$\times$10$^{-3}$ pc$^{-3}$.  Assuming a flat detection limit between 0\farcs{}25 and 1\farcs{}1 corresponding to an object at 10$^{-6}$ contrast and adopting a conversion to absolute $H$ band magnitudes  \cite{Pecaut2013}, a background L7 (L5, T0) dwarf must be at a distance closer than $\sim$179 $pc$ ($\sim$245 $pc$, $\sim$144 $pc$) to be detected.   Multiplying by the space density for L5--T0 dwarfs, we determine the contamination probability of an L5--T0 dwarf is $\approx$3$\times$10$^{-7}$.  Considering the total number of accelerating stars observed observed with SCExAO to date ($\sim$50), the survey-wide contamination rate is $\sim$1.5$\times$10$^{-5}$.

Third, HIP 99770's astrometric acceleration is consistent with that from a companion with the mass and separation we find for HIP 99770 b.  It an interloping background star is unlikely to have those properties.  Of the $\sim$10 other candidate companions imaged within SCExAO/CHARIS field from our accelerating star survey to date, none have later been shown to be background stars. 

In summary, for HIP 99770 b to be an interloping field object, it would have to i) have a high non-zero proper motion that matches the system's proper motion and apparent orbital motion \textit{and} ii) be a late-type field L dwarf.   Considered separately, the probabilities of HIP 99770 b having these properties are $<$6.5$\times$10$^{-9}$ and 1.5$\times$10$^{-5}$.  Assuming they are independent, the joint probability is then $<$1$\times$10$^{-13}$, or less than one in ten trillion ($>$7.4-$\sigma$).   Additionally, a background substellar object would have to be located at a position that is consistent with HIP 99770's proper motion anomaly.  

We therefore reject the possibility that HIP 99770 b is a background object.

\textbf{Astrometric Modeling}: 
The joint modeling of astrometry and direct imaging data was required to discover HIP 99770 b and identify it as a planet.  Astrometry alone cannot yield the mass or orbit of HIP 99770 b.   
The secondary companion's mass and orbit cannot be directly determined from the star's proper motion anomaly for companion orbital periods longer than the 25-year time baseline established by Hipparcos and Gaia.  The acceleration, $a_{\rm PM}$, is degenerate with the companion mass, $M_{\rm sec}$, and the angular separation, $\rho$, so $M_{\rm sec}$ $\propto$ $a_{\rm PM}\rho^{2}$ (\cite{Brandt2019}, their equations 4-6). 
In other words, a companion inferred through Hipparcos and Gaia astrometry could either be a low-mass companion at small separations or a more massive one at wide separations. 

The astrometric acceleration for a given companion mass depends on $\phi$, the angle between the  object’s position vector and the plane of the sky (\cite{Brandt2019}, their equation 4).   Astrometry does not constrain $\phi$.  As the primary is an A-type star with very weak metal lines, useful RV constraints on $\phi$ are not available.    Direct imaging can break the degeneracy, by determining  $\rho$.   Jointly modeling relative astrometry from imaging with the star’s absolute astrometry can then determine $\phi$.

Conversely, direct imaging alone would not conclusively determine whether HIP 99770 b is a planet.   Luminosity evolution models used to infer mass are uncertain due to the lack of observational constraints on their key assumptions \cite{Berardo2017,Spiegel2012}.  For HIP 99770 b, this uncertainty would be exacerbated because the system's age is poorly constrained.  Joint modelling of both observational datasets is required to determine the companion's mass.   

Figures S8 and S9 compare the absolute astrometry of the HIP 99770 primary and relative astrometry of HIP 99770 b to model values drawn from the posterior probability distribution of orbits from {\tt orvara}.  
They confirm that HIP 99770 is accelerating, as evidenced by the difference in $\mu_{\rm \alpha}$ between Hipparcos and Gaia.   Comparing these two measurements to the integral constant -- the proper motion based on the Hipparcos-Gaia scaled positional difference (Table S1) -- also indicates an acceleration.

Figures S8 and S9 also show that the {\tt orvara} posteriors match the absolute astrometry for the HIP 99770 primary and relative astrometry for HIP 99770 b.  The $\chi^{2}$ values for the best-fitting posterior draw are $\chi^{2}$ $\sim$ 0.015, 0.52, and 2.79 for the Hipparcos-Gaia scaled positional difference measurements, the Gaia measurements, and the Hipparcos measurements, respectively.   For HIP 99770 b's angular separation and position angle, the $\chi^{2}$ values are 0.12 and 1.41, respectively.  Residuals for the Hipparcos astrometry differ by $\approx$1-$\sigma$, and residuals between the {\tt orvara} posteriors and all other measurements are smaller.

HIP 99770 b is unlikely to be a brown dwarf with a mass greater than 25 $M_{\rm J}$.  Figure S8A and Figure S8B show that increasing the mass to 25--40 $M_{\rm J}$ moves the proper motion anomaly for Hipparcos in the wrong direction (i.e. $\mu_{\rm \alpha}$ (Hipparcos) $\sim$ 70--72 $mas$ yr$^{-1}$), predicts too small of a proper motion for Gaia, and produces a nearly constant angular separation, which are all inconsistent with the data.   

  We explored the sensitivity our results to input priors for the stellar mass and planet mass (Table S5).   For the assumed stellar mass, we varied the peak and standard deviation for our primary mass prior.  For the planet mass, we tested a uniform prior (our nominal case) and a 1/$M_{\rm p}$ prior.   
Table S6 lists our results.   The covariances between the stellar mass and planet mass are weak.   Increasing (decreasing) the prior peak value from 1.8 $M_{\rm \odot}$ to 1.9 $M_{\rm \odot}$ (1.7 $M_{\rm \odot}$) results in only a 3\% increase (4.4\% decrease) in the median of the planet mass posterior distribution.   Changing the width of the prior distribution has no effect on the planet mass.  Adopting a 1/$M_{\rm p}$ prior on the planet mass leads to a posterior with a median value smaller by $\sim$ 15\%: 13.9 $M_{\rm J}$.  This shift is less than the 1-$\sigma$ uncertainty on the inferred mass.

We also investigated the log likelihood ($ln$ $\Lb$) of the orbit as a function of companion mass drawn from our posterior probability distribution.  The log likelihood peaks at $\approx$ 15--17 $M_{\rm J}$, and the upper envelope of the distribution drops by $\Delta ln \Lb \approx$4 at 4 $M_{\rm J}$ and 30 $M_{\rm J}$: orbits with a companion mass at the median of the posterior distribution are the best fitting to the data.  Finally, we re-ran our analysis after doubling the astrometric errors in each epoch, effectively making the CHARIS astrometry uncertain to within 50-60\% of a pixel in each coordinate for each epoch and the NIRC2 astrometry uncertain by 1.2 pixels in each coordinate.   These uncertainties are larger than those we empirically estimated by comparing astrometry determined from different reductions for each data set (e.g. ADI vs ASDI reductions of CHARIS data: $\sim$ 2--3 mas).  The resulting companion mass is 17.5$^{+8.7}_{-5.4}$ $M_{\rm J}$ -- less precise but consistent with our fiducial results.
 
 \textbf{Spectral Analysis and Atmospheric Modeling}: HIP 99770 b's near-IR colors provide information on the object's atmosphere.
Figure S10 shows that HIP 99770 b's has properties close to the L/T transition, similar to the location of young low-gravity brown dwarfs.   HR 8799 cde and the dusty, cloudy planet-mass companion VHS J125601.92–125723.9 B (hereafter VHS 1256-1257 B) appear slightly redder \cite{Marois2008,Marois2010,Gauza2015}.    Figure S11 shows that an L7 spectral type template provides the best match to HIP 99770 b's spectrum.
 
 We then compared the planet's CHARIS $JHK$ spectra and NIRC2 $L_{\rm p}$ photometry to the library of empirical spectra from the Montreal Spectral Library and to HR 8799 d \cite{Gagne2014,Greenbaum2018,Zurlo2016}.   The empirical library covers field and young substellar objects ranging from early M to T spectral types.  We find that HIP 99770 b is best matched by objects at the L/T transition (Figure S12), specifically the L9.5 dwarf SIMPJ0956-1447.   HIP 99770 b has a spectrum intermediate in flatness between HR 8799 d and this L9.5 dwarf, indicating that its cloudiness and level of atmospheric dust is intermediate between a very young L/T transition exoplanet and a field brown dwarf.   
 HIP 99770 b's absolute magnitude in the $L_{\rm p}$ filter and its bolometric luminosity matches that of HR 8799 d as well as HR 8799 ce: $M_{\rm L_{\rm p}}$ $\sim$ 11.5--11.6 and log($L$/$L_{\odot}$) $\sim$ -4.7 $\pm$ 0.1 \cite{Currie2014b,Marley2012}.  However, HIP 99770 b and HR 8799 d differ by the flatness of their spectra, which probes clouds/dust, and their $K$ band spectral shapes, probing chemistry.
 
 
We compare HIP 99770 b's CHARIS spectrum and NIRC2 photometry to atmosphere models 
\cite{Burrows2006,Currie2011, Allard2012, Madhusudhan2011}.  We first consider the BT-Settl family of models.
The BT-Settl models include a self-consistent model for clouds and dust entrained in clouds; we adopt the Asplund et al. \cite{Asplund2009} abundances.  Our second set of models are the Lacy/Burrows models, which either parameterize clouds via a shape function that describes the cloud truncation relative to the scale height of the gas, or adopt cloud-free atmospheres.   From this model family we consider the cloud-free model grid and the following cloudy models (ordered from least cloudy to most cloudy): E, AEE, AE, and A models as defined in previous work \cite{Burrows2006,Madhusudhan2011}.  Following previous work \cite{Burrows2006,Madhusudhan2011}, we vary the modal size of silicate dust in the atmospheres, from 30 $\mu m$ to 100 $\mu m$.   

We updated the Lacy/Burrows models as follows.  First, we incorporated molecular line lists from the Exomol collaboration \cite{Tennyson2016,Chubb2021}, which substantially changed the pressure dependence of water opacity, which shapes the wings of HIP 99770 b's spectrum (see Figure 4A).   Pressure dependencies for other molecules -- e.g. metal hydrides, TiO -- were also updated \cite{Gharib-Nezhad2021}.  Absorption cross sections for the resonance doublets of K and Na perturbed by H$_2$ collisions are incorporated \cite{Allard2016,Allard2019}. Compared to the previous models, ours are flatter, more blackbody-like for a given cloud and dust prescription.

Second, we incorporate non-equilibrium carbon chemistry into the models \cite{Hubeny2007}.   Within this framework, the eddy-diffusivity value, K$_{zz}$ quantifies the strength of atmospheric mixing (see also \cite{Barman2011}).  We adopt a value of $K_{\rm zz}$ = 10$^{5}$ \cite{Hubeny2007}.  Comparisons between these models and our data showed stronger CH$_{\rm 4}$ absorption than was present in the data.   We therefore reduced the methane abundance by a factor of 10 (e.g. see \cite{Skemer2012} for a similar approach).   
Non-equilibrium chemistry produces slight changes in the wings of synthetic spectra at $H$ band and an increase in flux density at 2.2--2.4 $\mu m$.

As in prior work \cite{Currie2018a}, we focus only on the CHARIS channels unaffected by telluric absorption.
We quantify the goodness-of-fit for the $kth$ model using the $\chi^{2}$ statistic, considering the spectral covariance:
\begin{equation}
    \chi^{2} = R_{k}^{T}C^{-1}R_{k} + \sum_{i}(f_{phot,i}-\alpha_{k}~F_{phot,ik})^{2}/\sigma_{phot,i}^{2}.
\end{equation}
The vector $R_{k}$ is the difference between measured and predicted CHARIS data points ($f_{spec}-\alpha_{k}F_{spec}$), $R_{k^{T}}$ is its transpose, and $C$ is the covariance for the CHARIS spectra.  The vectors $f_{phot,i}$, $F_{phot,ik}$, and $\sigma_{phot,i}$ are measured photometry, model predicted photometry, and photometric uncertainty.  The scaling factor $\alpha_{k}$ is a free parameter we vary to minimize $\chi^{2}$ for a given model and is equal to the planet radius divided by the distance to HIP 99770.   We assume a distance of 40.74 $pc$.  

Cloudy models with gravities and temperatures of log(g) = 4--5 and 1250--1600 $K$ reproduce most of HIP 99770 b's spectrum and photometry (see main text).  Figure S13 compares HIP 99770 b's spectrum and photometry to to models assuming different cloud thicknesses and modal dust sizes.
The best-fitting models from the Lacy/Burrows suite come from two model grids.   The first grid assumes AEE-type clouds with a modal particle size of 100 $\mu m$.  The best-fitting temperature and gravity for models from this grid is 1300 K: and $log(g)$ = 4.5 ($\chi^{2}$ = 1.397): the 2-$\sigma$ confidence interval covers $T_{\rm eff}$ = 1250--1500 K and log(g) = 4--5.  The second grid assumes E-type clouds with a modal particle size of 60 $\mu m$.  The best-fitting temperature and gravity for models from this grid is 1400 $K$: and $log(g)$ = 4 ($\chi^{2}$ = 1.411): the 2-$\sigma$ confidence interval covers $T_{\rm eff}$ = 1350--1500 $K$ and log(g) = 4--4.5.  Models with large dust (100 $\mu m$) but thicker clouds (AE) fit the data more poorly.  Models with 60-$\mu m$ dust and AE-type, thick clouds (AE60) do not fit the data, producing too flat a spectrum.   The BT-Settl model grid has a best fit at 1500--1600 $K$ ($\chi^{2}_{\rm nu}$ = 1.270) and 2-$\sigma$ ranges of $T_{\rm eff}$ = 1400--1600 $K$ and log(g) = 4.   The Lacy/Burrows family of models that best reproduce the HR 8799 planet photometry and spectra assume thick clouds and moderately small dust sizes (A or AE cloud types; 60 $\mu m$ modal dust sizes).   Our modeling therefore favors atmospheres with thinner clouds than the HR 8799 planets, in agreement with our empirical analyses.

The AEE100 grid's best-fit model ($T_{\rm eff}$ = 1300 K, $log(g)$ = 4.5) has an implied radius of 1.05 $R_{\rm J}$, consistent with evolution model predictions of $\sim$ 1.1--1.2 $R_{\rm J}$ \cite{Baraffe2003,Spiegel2012}.  Best-fitting models from most other grids imply radii that are $\sim$ 17--30\% smaller than those predicted for young gas giants from evolutionary models \cite{Baraffe2003,Spiegel2012}, though they are admissible on physical grounds, as they are still large enough to be consistent with an electron degeneracy pressure-supported object).  However, some nominally well-fitting models -- especially those in the BT-Settl grid with higher temperatures (1500-1600 K) -- have implied radii that may be unrealistically small.  Other directly-imaged planets show slight differences between radii derived from atmospheric modeling and radii expected from luminosity evolution models.  Even some state-of-the-art atmospheric retrievals applied to well studied planets -- e.g. HR 8799 e -- can show differences on the order of 10--20\%\cite{Molliere2020}, so differences between radii we derive for HIP 99770 b from atmospheric modeling and those expected from evolutionary models are not unusual.   


\textbf{HIP 99770 b's Luminosity}: Given the possible ages of HIP 99770 b (40 $Myr$, 115--414 $Myr$) and HIP 99770 b's luminosity of log($L$/$L_{\rm \odot}$) = -4.53 $\pm$ 0.02, the Baraffe hot start evolutionary models imply a mass of either $\sim$9 $M_{\rm J}$ or 11--32.5 $M_{\rm J}$ \cite{Baraffe2003}.    Alternate models -- \cite{Burrows1997,Marley2021} -- yield similar ranges: 11 $M_{\rm J}$ or 14-32 $M_{\rm J}$ and 9.5 $M_{\rm J}$ or 11.5--30 $M_{\rm J}$, respectively.  HIP 99770 b's range of radii and gravities from best-fitting models (shown in Figure 4)implies masses between 2.4 $M_{\rm J}$ and 39.5 $M_{\rm J}$.  Thus, direct dynamical mass measurements drawn from a small number of astrometric data points constrain HIP 99770 b's mass with far greater precision than possible with either luminosity evolution models or atmospheric models.

HIP 99770 b's luminosity is inconsistent with cold start luminosity evolution, gas giant planets formed with a low initial entropy, but is consistent with hot start, high-entropy predictions \cite{Spiegel2012}.  If hot start models -- \cite{Baraffe2003} -- are assumed, HIP 99770 b's luminosity and dynamical mass favor an age between 80 $Myr$ and 200 $Myr$.   Given our age estimate for the host star, the most likely age for HIP 99770 b would then be 115--200 $Myr$, which is consistent with the age derived above astroseismology if the star is a rapid rotator. 

\textbf{Interpreting HIP 99770 b}: 
HIP 99770 b's detection differs from most prior joint detections of companions from direct and indirect techniques, where RV detections are used in place of astrometry.
The earliest such joint efforts -- e.g. the TRENDS survey -- focused on direct imaging follow-up of stars with long-term RV trends indicative of a long-period companion \cite{Crepp2012}.  TRENDS found multiple M dwarf companions, a white dwarf, and a 65 Jupiter-mass, high-eccentricity brown dwarf: no planets or low-mass brown dwarfs \cite{Gonzales2020,Crepp2016}.   

Other efforts targeting stars with long-term RV trends have imaged stellar companions or high-mass and high eccentricity brown dwarfs but not planets \cite{Cheetham2018}.  Most substellar companions detected with both RV and imaging are brown dwarfs imaged well after their discoveries or vice versa \cite{Rickman2019,Rickman2020, Bowler2018}.   In one special case thus far, the nearby star $\beta$ Pic, one imaged planet ($\beta$ Pic b) was detected with RV a decade later after an intense monitoring campaign and a second planet ($\beta$ Pic c) was discovered with RV and then detected later with high-contrast interferometry \cite{Lagrange2010,Lagrange2019,Lagrange2020,Nowak2020}.  
 HIP 99770 b was detected through the combination of direct imaging and precision astrometry.   Most prior attempts using astrometry to discover planets around main sequence stars have led to claims later refuted, null detections, unconfirmed candidates, or brown dwarf/M dwarf companions instead \cite{Gatewood1973,Bean2010,Muterspaugh2010,Sahlmann2014}.   Previous exoplanets with astrometric detections besides HIP 99770 b had already been detected using other methods much earlier (e.g. \cite{Venner2021,Brandt2021c}) or are still candidates \cite{Curiel2020}.
 
 Figure 4 compares HIP 99770 b's mass ratio and separation to that of other directly imaged objects and RV-detected planets.   For planets, we use the NASA Exoplanet Archive database (https://exoplanetarchive.ipac.caltech.edu/).   The exoplanet.eu webpage catalogues objects up to 60 $M_{\rm J}$ as possible planets (or brown dwarfs) \cite{Schneider2011}.  
 
We combine the two data sources and make several modifications and corrections.  We revise upward the masses for imaged companions 1RXJ1609.1-210524 B, GJ 504 B, GSC 06214-00210 B, HIP 78530 B, and HR 7329 B based on updated system ages \cite{Pecaut2012,Bonnefoy2018,Mamajek2014}.   Similarly, we revise the mass for VHS 1256 B and $\kappa$ And b  \cite{Dupuy2020,Currie2018a}.  For the HR 8799 planets, we adopt the dynamical mass of HR 8799 e of 9.6$^{+1.9}_{-1.8}$ $M_{\rm J}$ as the mass for HR 8799 cd \cite{Brandt2021c} and set the mass of HR 8799 b to be 6.3$^{+1.4}_{-1.3}$ $M_{\rm J}$ as a compromise between several proposed \cite{Marois2008,Marois2010,Currie2011,Wang2018}.  HD 206893 b is added to our sample despite it not appearing in the Archive \cite{Milli2017}.   Similarly, we add the protoplanet AB Aur b \cite{Currie2022}.   The Archive's listed mass for HD 100546 b is revised to literature values \cite{Currie2015}.  We remove LkCa 15 bc from our sample since these candidate protoplanets have been shown to be  refuted \cite{Currie2019}.   Finally, we revert the mass estimate of BD+20 2457 b to 21.4 $M_{\rm J}$ \cite{Niedzielski2009}.
 
The population of exoplanets detected using RVs has an upper envelope of $q$ $\sim$ 0.01 from 1 to $\sim$ 30 au, with a slightly smaller number of companions with higher mass ratios ($q$ $\sim$ 0.01--0.025).   Some RV-detected planets with discrepant mass ratios have large uncertainties (e.g. HD 66428 b).   Multiple RV-detected planets have similar mass ratios/separations to HIP 99770 b.  The RV sample has low completeness at $a_{\rm p}$ $>$ 10 au \cite{Rosenthal2021}.  The imaged brown dwarf companions HIP 75056 AB, HIP 74865 B, and HR 2562 B and planet/brown dwarf companion GJ 504 B have $q$ $\sim$ 0.015--0.025 \cite{Wagner2020,Hinkley2015,Kuzuhara2013,Konopacky2016} : i.e. at or slightly below the turnover in the mass ratio distribution of substellar companions \cite{Reggiani2016}, although some of these companions (e.g. HR 2562 B) have masses that are not well constrained as they are highly sensitive to the assumed system age and/or adopted luminosity evolution model.   Other than those targets, there is a lower limit of $q$ $\sim$ 0.025 for objects unanimously classified as brown dwarfs.  The lowest any object more massive than 25 $M_{\rm J}$ extends is $q$ $\sim$ 0.015.   HIP 99770 b has $q$ $\sim$ 0.0084 and $a_{\rm p}$ $\sim$ 16.9 au, so it lies well within the population defined by bona fide planets.

 HIP 99770 b could plausibly have formed in a protoplanetary disk.  At a semimajor axis of $\approx$17 au, it orbits at a separation that is small compared to the physical size of most protoplanetary disks and its companion-to-primary mass ratio is smaller than typical ratios of protoplanetary disk masses to primary masses.  
  Protoplanetary disk radii span a range of values but their distributions peak at $\sim$200 au and fall to low frequencies by $\sim$300 au \cite{Andrews2007}.  While absolute values for disk masses rely on the gas-to-dust ratio and thus are uncertain by a factor of several, adopting solar values yields a median disk-to-star mass ratio $q_{disk}$ $\sim$ 0.01, where almost all disks have $q_{disk}$ $<$ 0.1 \cite{WilliamsCieza2011}.  Binary companions to more massive objects with mass ratios of $q$ $\le$ 0.05 are rare \cite{Kraus2008}; the distribution of substellar mass ratios turns over at $q$ $\sim$ 0.025 \cite{Reggiani2016}.  Imaged objects with mass ratios of $q$ $\le$ 0.01--0.025 and separations of $a_{\rm p}$ $\le$ 200--300 au roughly define a population contiguous with that of planets detected through indirect means. 
 
\textbf{Further Evidence that HIP 99770 b is a Planet, Not a Brown Dwarf}: Like $\kappa$ And b \cite{Currie2018a}, HIP 99770 b's likely mass is near the deuterium-burning limit, roughly 13 $M_{\rm J}$. Many early studies identify planets as objects with masses below the deuterium-burning limit based on criteria suggested in 2003 by the \textit{International Astronomical Union}'s \textit{Working Group on Extrasolar Planets} (WGEP) \cite{Boss2007}.  WGEP explicitly notes that this definition is not a normative statement, is focused on interpreting detections around mostly Sun-like stars and free-floating objects only, and can evolve as more companions are detected around different types of stars.

Distinguishing between a gas giant planet and a brown dwarf based on deuterium burning is 
often rejected in practice within the direct imaging community and in major exoplanet catalogues such as the NASA Exoplanet Archive and exoplanet.eu.   
 Deuterium burning is time, metallicity, and helium abundance dependent \cite{Spiegel2011}; unlike hydrogen burning, it does not identify a meaningful boundary for the evolution of low-mass objects at all \cite{Chabrier2007,Luhman2008}.    
During first $\sim$5--10 Myr, the critical time for setting the final mass of a jovian planet, the luminosities of objects up to 20 $M_{\rm J}$ are still dominated by Kelvin-Helmholtz contraction  \cite{Spiegel2011}: it is extremely difficult to see how the ignition of deuterium could affect nebular gas accretion, let alone shut it off the moment 13 $M_{\rm J}$ of material is accreted.   Objects in the $\sim$ 15 $M_{\rm J}$ range, comparable to HIP 99770 b's best-estimated mass, do not burn deuterium at all for the first tens of Myr after formation \cite{Spiegel2011}, potentially implying that an object could literally transform from a planet into a brown dwarf well after it formed.  

On the observational side, recent imaging surveys have identified objects that are members of quadruple systems, clearly formed by molecular cloud fragmentation (i.e. like stars), with inferred masses down to 5 $M_{\rm J}$: some free-floating objects have sub-deuterium burning masses as well \cite{Todorov2010,Liu2013}.   RV surveys have identified some systems -- e.g. the 2.7 $M_{\odot}$ star  $\nu$ Oph -- with companions at $\sim$ 1 au with masses of 22 and 24 $M_{\rm J}$ ($q$ $\sim$ 0.008--0.009) that are nevertheless locked in a mean-motion resonance indicating formation in a disk (i.e. as a planet) \cite{Quirrenbach2019}.

Below we augment the analysis presented in Figure 4 that instead favor 
 demographics-driven diagnostics presumably connected to formation processes, investigating the substellar companion mass function 
 from surveys sensitive to both superjovian planets and brown dwarfs \cite{Sahlmann2011,Rosenthal2021}. 
Previous literature on the companion mass function for RV-detected companions with orbital periods less than $\sim$several years reaches a local minimum at minimum masses of $m~sin(i)$ $\sim$ 20--30 $M_{\rm J}$, not 13 $M_{\rm J}$ \cite{Grether2006,Udry2010,Sahlmann2011,Kiefer2019}, where objects less (more) massive than this limit are contiguous with a distribution drawn from Jupiter-mass (brown dwarf) companions.   As these are lower mass limits, the true minimum occurs at slightly larger masses.   Furthermore, the minimum may be proportional to the primary mass, indicating that companion mass ratio -- not mass -- is a more fundamental discriminator \cite{Grether2006}.  

To provide a modern empirical analysis of the planet and brown dwarf mass trends, we analyzed data drawn from the California Legacy Survey (CLS), an RV-survey with sensitivity to superjovian-mass planets out to $\sim$ 10 au \cite{Rosenthal2021}.   As shown in Figure S14, the CLS survey data supports a minimum in the companion mass function of $m~sin(i)$ $\sim$ 16--25 $M_{\rm J}$.   Given the average expected line-of-sight inclination to these systems, the minimum in absolute mass is $\sim$25--39 $M_{\rm J}$, about two to three times the mass of the deuterium-burning limit and higher than HIP 99770 b's dynamical mass.  Earlier work based on smaller samples also arrives at similar conclusions \cite{Sahlmann2011,Ma2014}.

 


An object's orbital eccentricity may also diagnose formation mechanisms \cite{Bowler2020}.   Very low-mass planets like $\beta$ Pic b or HR 8799 bcde have low eccentricities.   Other companions -- including those well above 30 $M_{\rm J}$ and $q$ $\sim$ 0.02 -- have a broad range of eccentricities.   HIP 99770 b's low eccentricity is consistent with the planet population.   As stated in the main text, HIP 99770 b receives roughly as much light as does Jupiter from the Sun.

Considered holistically, HIP 99770 b's properties -- insolation and mass ratio/separation, mass, and eccentricity -- support interpreting it as a bona fide planet, not a brown dwarf.

\newpage

  \begin{table}[]
  \captionsetup{labelformat=empty}
    \caption{ Table S1: \textbf{Propreties of the HIP 99770 Host Star}.  We list intrinsic properties of the star drawn from the literature or derived from this work, proper motion derived from the Hipparcos and Gaia missions, and photometry derived from this work.}
    \begin{tabular*}{1.0\textwidth}{c}
   \hline
    \end{tabular*}
    \small
     \begin{tabular*}{1.0\textwidth}{p{0.28\textwidth} m{0.49\textwidth} m{0.15\textwidth}}
     Property & {Value} & References\\
     \hline
     $T_{\rm eff}$  & 8000 K & \cite{Murphy2017}, this work\\
     Spectral Type & kA1.5hA7mA0.5 $\lambda$ Boo (A5--A6) & \cite{Murphy2017}, this work  \\
     Mass & 1.85 $\pm$ 0.19 $M_{\odot}$ & \cite{Jones2016}, this work\\
     Rotation Rate (Projected) & 65 $km$ $s^{-1}$ & \cite{Royer2007}\\
     Apparent Luminosity  & 13.86$^{+2.14}_{-5.46}$ $L_{\odot}$ & this work\\
     Age & 40 Myr, 115-414 Myr & this work\\
     Distance  & 40.74 $\pm$ 0.15 $pc$ & \cite{GaiaeDR32021}\\
     Proper~Motion (Gaia-eDR3) & 68.09 $\pm$ 0.12, 69.40 $\pm$ 0.14 mas yr$^{-1}$ & \cite{GaiaeDR32021,Brandt2021}\\
     Proper Motion (H-G scaled$^{\ast}$) & 68.24 $\pm$ 0.01, 69.67 $\pm$ 0.01 mas yr$^{-1}$& \cite{GaiaeDR32021,Brandt2021}\\
     Proper Motion (Hipparcos) & 69.45 $\pm$ 0.38, 69.19 $\pm$ 0.38 mas yr$^{-1}$& \cite{GaiaeDR32021,Brandt2021}\\
     Proper Motion Anomaly ($\chi^{2}$) & 7.23& \cite{Brandt2021}\\
     J,H,Ks,L$_{\rm p}$ (mag) & 4.49 $\pm$ 0.05, 4.46 $\pm$ 0.02, 4.42 $\pm$ 0.02, 4.40 $\pm$ 0.05 & this work \\
     \hline
     \end{tabular*}
     {\raggedright \scriptsize{$\ast$ - H-G scaled stands for the average proper motion between the Hipparcos and Gaia missions.  }}
\end{table}

\newpage
 \begin{table}[]
   \captionsetup{labelformat=empty}
  \caption{Table S2: \textbf{HIP 99770 Observing Log}.  The table lists the observing date, instrument, astronomical seeing at optical wavelengths, the passbands covered by each data set and their wavelength ranges, cumulative exposure times, and total parallactic angle rotations.}
      \begin{tabular*}{1.0\textwidth}{l|l|l|c|c|c|c}
      \hline
        UT Date & Instrument & Seeing ($^{\prime\prime}$)& Passband & $\lambda$ ($\mu m$)& $t_{\rm exp}$ (s) & $\Delta$PA ($^{o}$)\\
     \hline
     20200729 & SCExAO/CHARIS & 0.7-0.8 & $JHK$ & 1.16--2.37 & 1569 & 34.5 \\
     20200901 & SCExAO/CHARIS & 0.7-1.1$^{a}$ & $JHK$ & 1.16--2.37 & 2031 & 92.9  \\
     20210519 & SCExAO/CHARIS & 0.3-0.5$^{b}$ & $JHK$ & 1.16--2.37 & 2353 & 87.6\\
     20210603 & Keck/NIRC2 & 0.3-0.45 & $L_{\rm p}$ & 3.78 & 3750 &61.9\\
     20210713 & SCExAO/CHARIS & 0.8-1.0 & $JHK$ & 1.16--2.37 & 4337 & 62.5\\
     20211017 & SCExAO/CHARIS & N/A$^{c}$ & $JHK$ & 1.16--2.37 & 5041 & 75.6\\
    \hline
     \end{tabular*}
  {\raggedright \scriptsize{a) The original sequence was far longer: we removed 70\% of frames suffering from periodically poor AO corrections due to variable seeing. b) While conditions for AO were good, the observations suffered from Low-Wind Effect, leading us to remove 60\% of frames due to splitting/smearing of the PSF. c) No seeing estimate was available; raw SCExAO/CHARIS contrasts were intermediate between July 2021 and May 2021 values.} \par}
     \label{tab:obslog}
 \end{table}

  \begin{table}[]
   \captionsetup{labelformat=empty}
    \caption{ Table S3: \textbf{Astrometric Data for HIP 99770 b}.  We list the observing date, instrument, relative astrometry of HIP 99770 b with respect to its host star, and the signal-to-noise ratio of HIP 99770 b's detection for data processed with ADI or ADI+SDI (ASDI)}
    \small
      \begin{tabular*}{1.0\textwidth}{c}
     \hline
     \end{tabular*}
      \begin{tabular*}{1.0\textwidth}{p{0.1\textwidth} p{0.25\textwidth} m{0.35\textwidth} m{0.25\textwidth}}
  Date & Instrument & Position [E,N]\farcs{} & SNR (ADI, ASDI)\\
     \hline
    20200729 & SCExAO/CHARIS & [0.263,-0.367] $\pm$ [0.004,0.005] & 5.2, 11.3\\
      20200901 & SCExAO/CHARIS & [0.263,-0.366] $\pm$ [0.005,0.005]& 7.3, 16.3 \\
     20210519 & SCExAO/CHARIS &[0.280,-0.343] $\pm$ [0.004,0.004]  & 17.3, 39.4\\
     20210603 & Keck/NIRC2 &[0.286,-0.337] $\pm$ [0.006,0.006] & 11 \\
     20210713 & SCExAO/CHARIS & [0.286,-0.338] $\pm$ [0.004,0.004]& 11.7, 22.8\\
     20211017 & SCExAO/CHARIS & [0.292,-0.327] $\pm$ [0.004,0.004]& 24.9, 45.0\\
     \hline
    \end{tabular*}
\end{table}
 

  \begin{table}[]
  \captionsetup{labelformat=empty}
    \caption {Table S4: \textbf{Measured and Derived Properties of HIP 99770 b.}   Photometry in the $J$, $H$, and $K_{\rm s}$ passbands is in units of magnitudes and is determined by integrating the SCExAO/CHARIS spectrum over each filter's transmission profile.  Atmospheric properties are determined from atmospheric model fitting.  HIP 99770 b's mass, mass ratio, and orbital properties are determined from a joint modeling of HIP 99770's absolute astrometry and HIP 99770 b's relative astrometry.}
          \begin{tabular*}{1.0\textwidth}{c}
     \hline
     \end{tabular*}
      \begin{tabular*}{1.0\textwidth}{p{0.5\textwidth} p{0.5\textwidth}}
      m$_{J}$ (1.25 $\mu m$) & 17.39 $\pm$ 0.19\\
       m$_{H}$ (1.65 $\mu m$) & 16.51 $\pm$ 0.11 \\
       m$_{K_{\rm s}}$ (2.16 $\mu m$) & 15.66 $\pm$ 0.09\\
       m$_{L_{\rm p}}$ (3.78 $\mu m$) & 14.52 $\pm$ 0.12\\
       Spectral Type & L7-L9.5\\
       $T_{\rm eff}$  & 1400$^{+200}_{-150}$ K\\
       log$(L/L_{\rm \odot})$ & -4.53 $\pm$ 0.02\\
       log($g$)  & 4--5\\
       Mass & 16.1$^{+5.4}_{-5.0}$ M$_{\rm J}$$^{\ast}$ \\
       Mass Ratio (q) & $0.0084^{+0.0028}_{-0.0026}$$^{\ast}$\\
       a$_{\rm p}$ & $16.9^{+3.4}_{-1.9}$ au\\
       e & $0.25^{+0.14}_{-0.16}$\\
        i & ${148}_{-11}^{+13}$ degrees \\
       Orbital Period & $51.0^{+17}_{-7.8}$ years\\
      \hline
    \end{tabular*}
     {\raggedright \scriptsize{$\ast$ Adopting a 1/$M_{\rm p}$ prior, the mass and mass ratio change to $M_{\rm p}$ = 13.9$^{+6.1}_{-5.1}$ M$_{\rm J}$ and $q$ = 0.0073$^{+0.0032}_{-0.0027}$.} \par}
\end{table}
 
    \begin{table}[]
    \captionsetup{labelformat=empty}
  \caption{Table S5: \textbf{Astrometric Priors for our} \texttt{orvara} \textbf{simulations}.  $\mathcal{N(\mu, \sigma)}$ represents a Gaussian with mean $\mu$ and variance $\sigma^2$: e.g. the mean and variance for the parallax is equal to the measured Gaia eDR3 parallax and 1-$\sigma$ uncertainty.  The variables $\omega$ and $\Omega$ are the argument of periastron and the position angle of the ascending node, respectively.   }
    \begin{tabular*}{0.9\textwidth}{l|c}
     \hline
     \hline
     \hspace{1cm} Parameter& Prior\\
     \hline
 
$M_{\rm{\star}}$ ($M_{\rm{\odot}}$) & $\mathcal{N}(1.8, 0.2)$\\
$M_{\rm{p}}$ ($M_{\rm{Jup}}$) & uniform (0--10$^{6}$)\\
a (au) &  $1/a$ (log-flat; a = 0 to 1 $pc$)\\
$\sqrt{e}\sin{\omega}$ & uniform (0--1)\\
$\sqrt{e}\cos{\omega}$ & uniform (0--1)\\
Inclination ($^{o}$) &  $\sin{i}$ ($\sin{i}$ = 0--1)\\
$\Omega$ ($^{o}$)   & uniform (0--2$\pi$)\\
Mean longitude at 2010.0 ($^{o}$) & uniform (0--2$\pi$)\\
Parallax (mas) & $\mathcal{N}(\varpi_{\rm{Gaia}}, \sigma_{\varpi,\rm{Gaia}})$ \\
     \hline
     \end{tabular*}
 \end{table} 
  
\newpage
 \begin{table}[]
    \captionsetup{labelformat=empty}
  \caption{Table S6: \textbf{Astrometric Modeling Tests using different priors for \texttt{orvara}}.  The first row -- highlighted in boldface font -- is our fiducial simulation, while other rows are simulations that explore the effect of changing our priors on the host star mass or companion mass.}
     \centering
    \begin{tabular*}{0.88\textwidth}{c|c}
     \hline
     \hline
     \hspace{2.825cm} Priors & Posteriors\\
     \hline
     \end{tabular*}
    \begin{tabular*}{0.82\textwidth}{l l|l l l l l}
        $M_{\rm \star}$ (M$_{\rm \odot}$) & $M_{\rm p}$ ($M_{\rm J}$) & $M_{\rm \star}$ (M$_{\rm \odot}$) & $M_{\rm p}$ ($M_{\rm J}$) & $a_{\rm p}$ (au) & $i_{\rm p}$ ($^{o})$  & $e_{\rm p}$ \\
     \hline
     \textbf{1.8 $\pm$ 0.2}  & \textbf{Uniform} & \textbf{1.85$^{+0.19}_{-0.18}$} & \textbf{16.1$^{+5.4}_{-5.0}$} & \textbf{16.9$^{+3.4}_{-1.9}$} & \textbf{148$^{+13}_{-11}$} & \textbf{0.25$^{+0.14}_{-0.16}$}\\
      1.8 $\pm$ 0.1  & Uniform & 1.81$^{+0.10}_{-0.10}$ & 16.1$^{+5.6}_{-5.0}$ & 16.7$^{+3.4}_{-1.7}$ & 149$^{+12}_{-11}$ & 0.26$^{+0.14}_{-0.16}$\\
      1.8 $\pm$ 0.1  & (1/$M_{\rm p}$) & 1.81$^{+0.10}_{-0.10}$ & 13.9$^{+6.1}_{-5.1}$ & 16.5$^{+3.6}_{-1.6}$ & 151$^{+12}_{-12}$ & 0.26$^{+0.13}_{-0.16}$\\
      1.7 $\pm$ 0.1  & Uniform   & 1.72$^{+0.10}_{-0.10}$ & 15.4$^{+5.8}_{-4.6}$ & 16.6$^{+4.0}_{-1.9}$ & 151$^{+13}_{-12}$ & 0.26$^{+0.14}_{-0.15}$\\
      1.9 $\pm$ 0.1  & Uniform & 1.90$^{+0.10}_{-0.10}$ & 16.6$^{+5.8}_{-4.9}$ & 16.9$^{+3.1}_{-1.7}$ & 146$^{+11}_{-10}$ & 0.24$^{+0.16}_{-0.16}$\\
    \hline
     \end{tabular*}
      \captionsetup{labelformat=empty}
     \label{tab:atmostests}
 \end{table} 
 
 \newpage
   \begin{figure*}
    \includegraphics[width=0.9\textwidth]{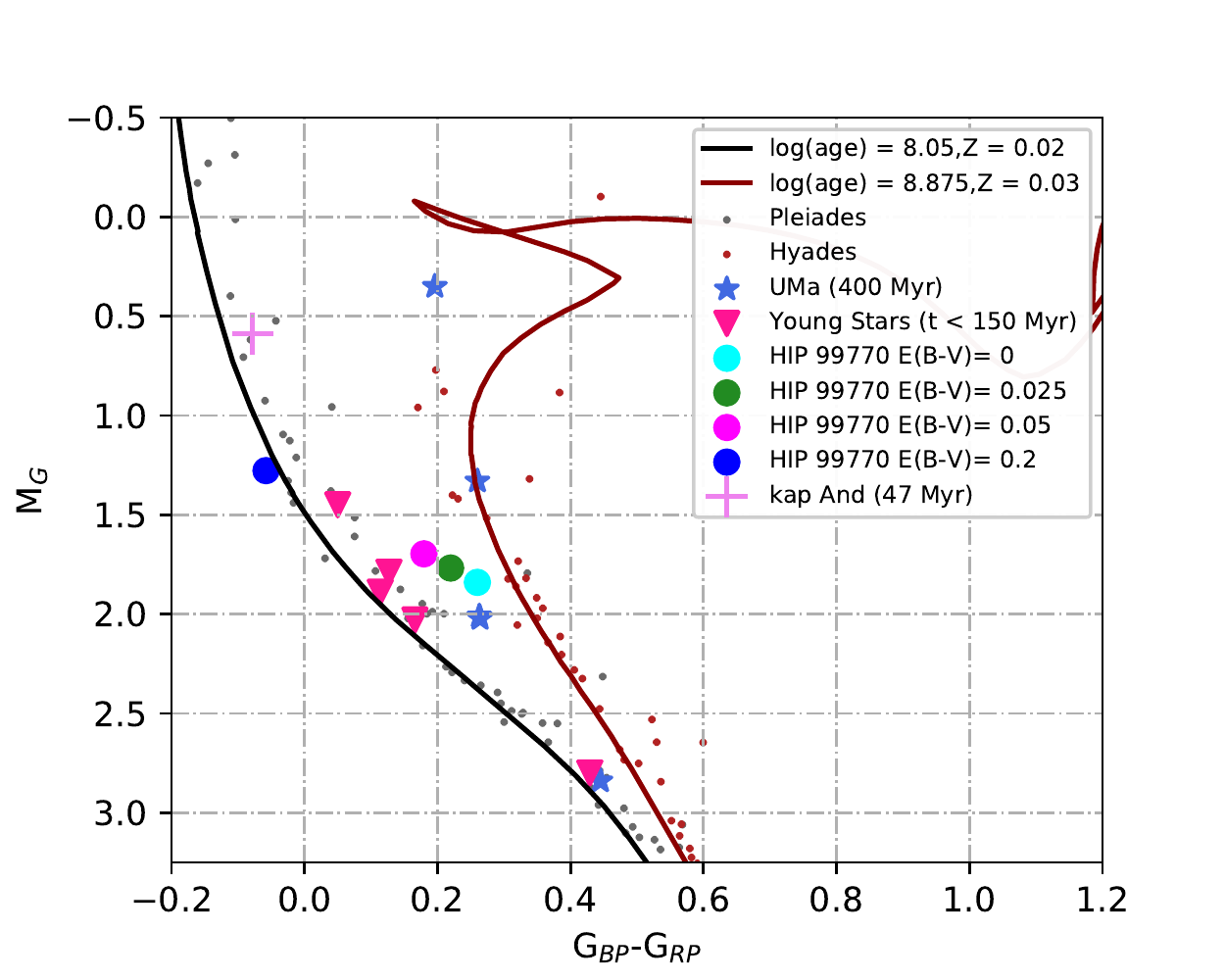} 
   \vspace{-0.2in}
   \captionsetup{labelformat=empty}
  \caption{Figure S1:
\textbf{Gaia color-magnitude diagram comparing the positions of stars with different ages to HIP 99770}.  We display members of the Pleiades ($\sim$115 Myr; grey dots) and Hyades ($\sim$750 Myr; red dots) \cite{Gaia2018}, isochrones for the Pleiades and Hyades \cite{Bressan2012}, select Ursa Majoris stars ($\sim$414 Myr; blue stars) with Center for High Angular Resolution Astronomy (CHARA) array data \cite{Jones2016}, and young stars with CHARA-estimated radii and ages (pink inverted triangles) \cite{Jones2016}.   We overplot isochrones that match the Pleiades locus (black line; 112 Myr, a metallicity of Z = 0.02) and the Hyades (dark red line; 750 Myr, Z = 0.03).   HIP 99770's measured absolute magnitude in the green Gaia passband is $M_{\rm G}$ = 1.84.  We also show color-magnitude diagram positions for HIP 99770 under different assumptions about its line-of-sight reddening.   If HIP 99770 is viewed nearly pole on, its apparent luminosity can be larger than its bolometric luminosity. 
  }
    \vspace{-0.15in}
\end{figure*} 

\begin{figure}
\begin{center}
\includegraphics[width=0.9\textwidth]{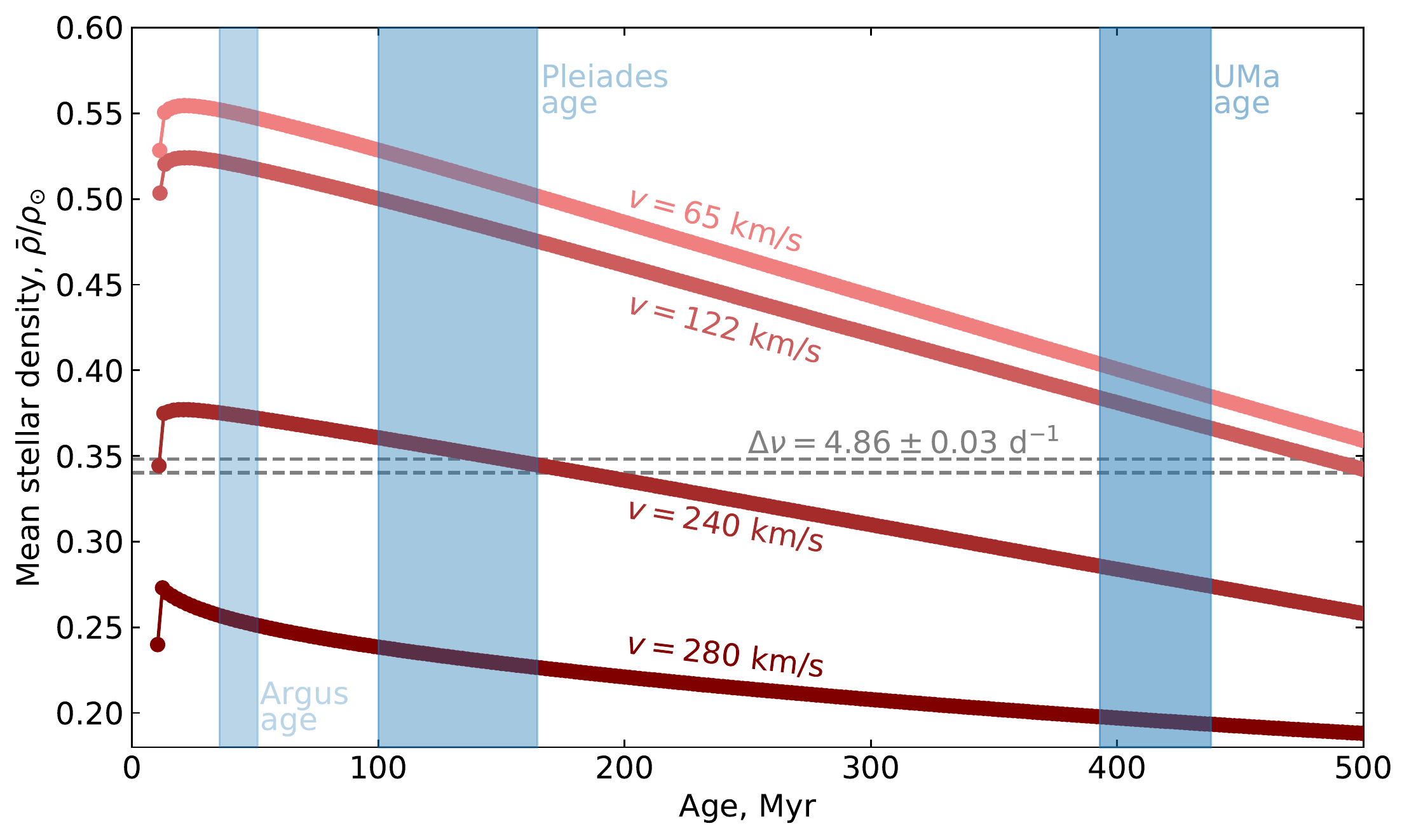}
\captionsetup{labelformat=empty}
\caption{Figure S2: \textbf{Mean stellar densities as a function of age for four different rotation rates}.  Red circles represent evolutionary tracks of four different stellar rotation rates (labeled), sampled every 2\,Myr from the end of the pre-MS ($\sim10.5$\,Myr) to 500\,Myr. Age ranges are shown for the Argus association (40 Myr; \cite{Zuckerman2019}), the Pleiades cluster ($\sim$115\,Myr; see discussion in \cite{murphyetal2022a}) and the Ursa Majoris Moving Group ($414\pm23$\,Myr; \cite{Jones2015}). The dashed grey line shows pulsation-derived density of $\bar{\rho}/\rho_{\odot}=0.34$, corresponding to $\Delta\nu = 4.86$.}
\end{center}
\end{figure}

\begin{figure*}
    \includegraphics[width=0.9\textwidth]{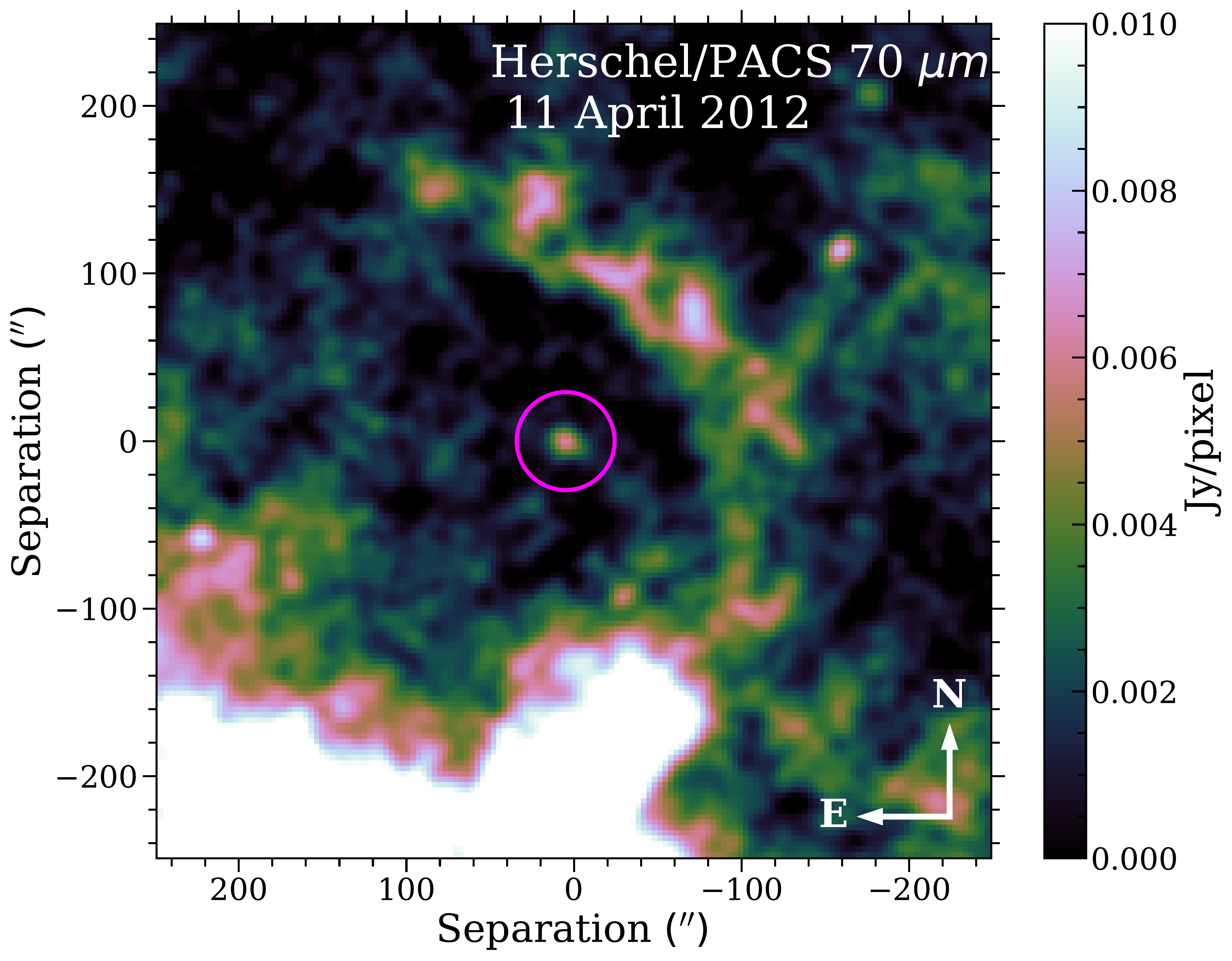} 
   \vspace{-0.1in}
   \captionsetup{labelformat=empty}
  \caption{Figure S3: \textbf{Detection of the HIP 99770 debris disk in archival Herschel 70 $\mu m$ data obtained with the Photodetector Array Camera and Spectrometer (PACS) instrument.}  The magenta circle shows the location of the star: the color bar is in units of Jy per pixel.   Background nebulosity surrounds HIP 99770 at an angular distance of roughly 50--100$\farcs{}$
  }
    \vspace{-0.5in}
\end{figure*} 
 
  \begin{figure*}
    \includegraphics[width=1.0\textwidth]{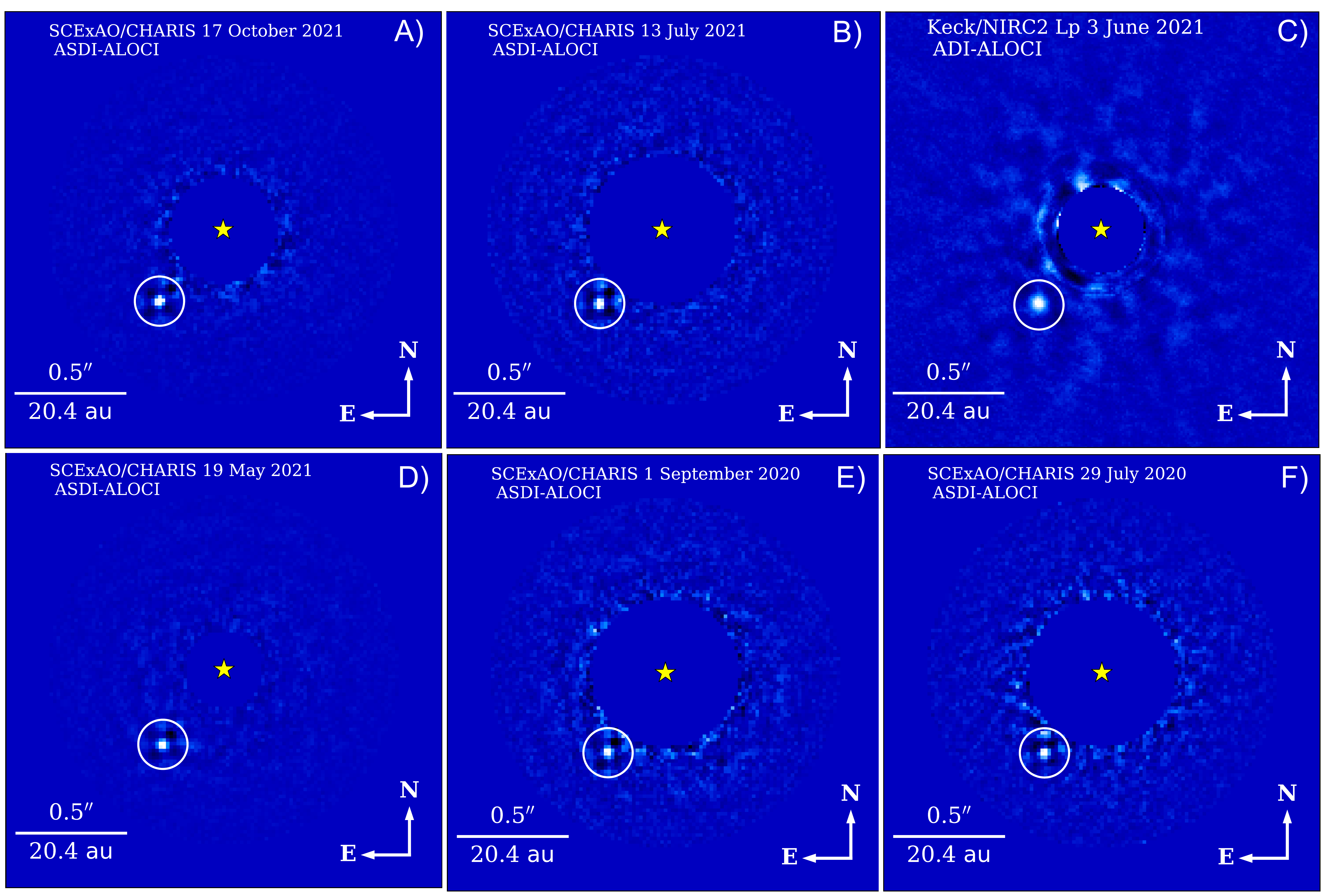} 
   \vspace{-0.3in}
   \captionsetup{labelformat=empty}
  \caption{Figure S4:
  \textbf{HIP 99770 images for each epoch.}   The meaning of the plot labels and the white circle, as well as the intensity scaling is the same as in Figure 1. 
  }
    \vspace{-0.5in}
\end{figure*} 

 \begin{figure*}[h!]
    \includegraphics[width=0.335\textwidth]{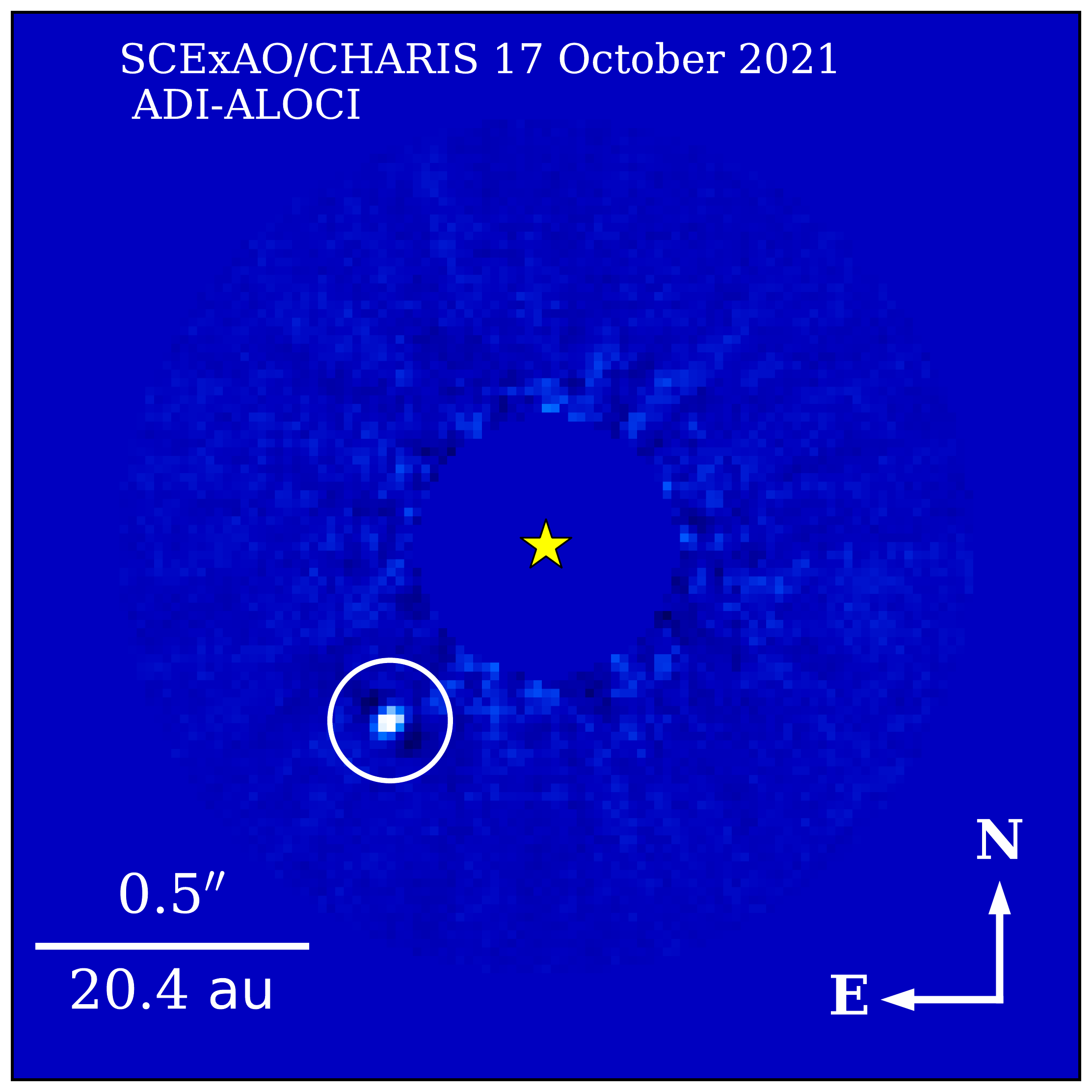} 
   \vspace{-0.1in}
   \captionsetup{labelformat=empty}
  \caption{Figure S5: \textbf{The 17 October 2021 SCExAO/CHARIS data reduced with ADI only}.  We use this reduction to extract HIP 99770 b's spectrum. 
  }
    \vspace{-0.0in}
\end{figure*} 

  \begin{figure*}
    \includegraphics[width=0.8\textwidth]{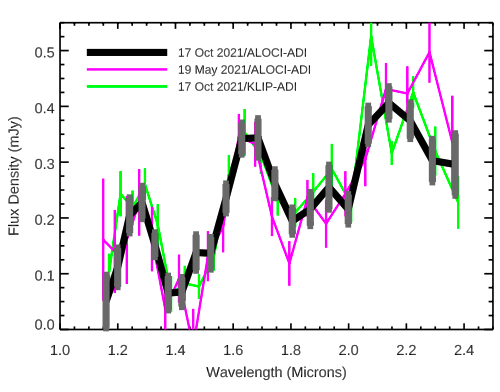} 
   \vspace{-0.15in}
   \captionsetup{labelformat=empty}
  \caption{Figure S6:
  \textbf{Spectra extracted for HIP 99770 b in different epochs and with different algorithms}.  We compare our adopted HIP 99770 b spectrum (October 2021, ALOCI-ADI; black line with gray error bars) with spectra extracted from May 2021 with the same algorithm (magenta line and error bars, offset to the left in wavelength) or with a different PSF subtraction algorithm (KLIP-ADI; green line and error bars, offset to the right in wavelength).   All error bars represent 1-$\sigma$ uncertainties.  The spectrum from May 2021 and the October 2021 spectrum obtained with KLIP-ADI both agree within errors to our adopted spectrum in 18 of the 21 spectral channels.
  }
    \vspace{-0.5in}
\end{figure*} 

 \begin{figure*}[h!]
  \centering
  \includegraphics[width=0.95\textwidth,trim=0mm 0mm 0mm 0mm,clip]{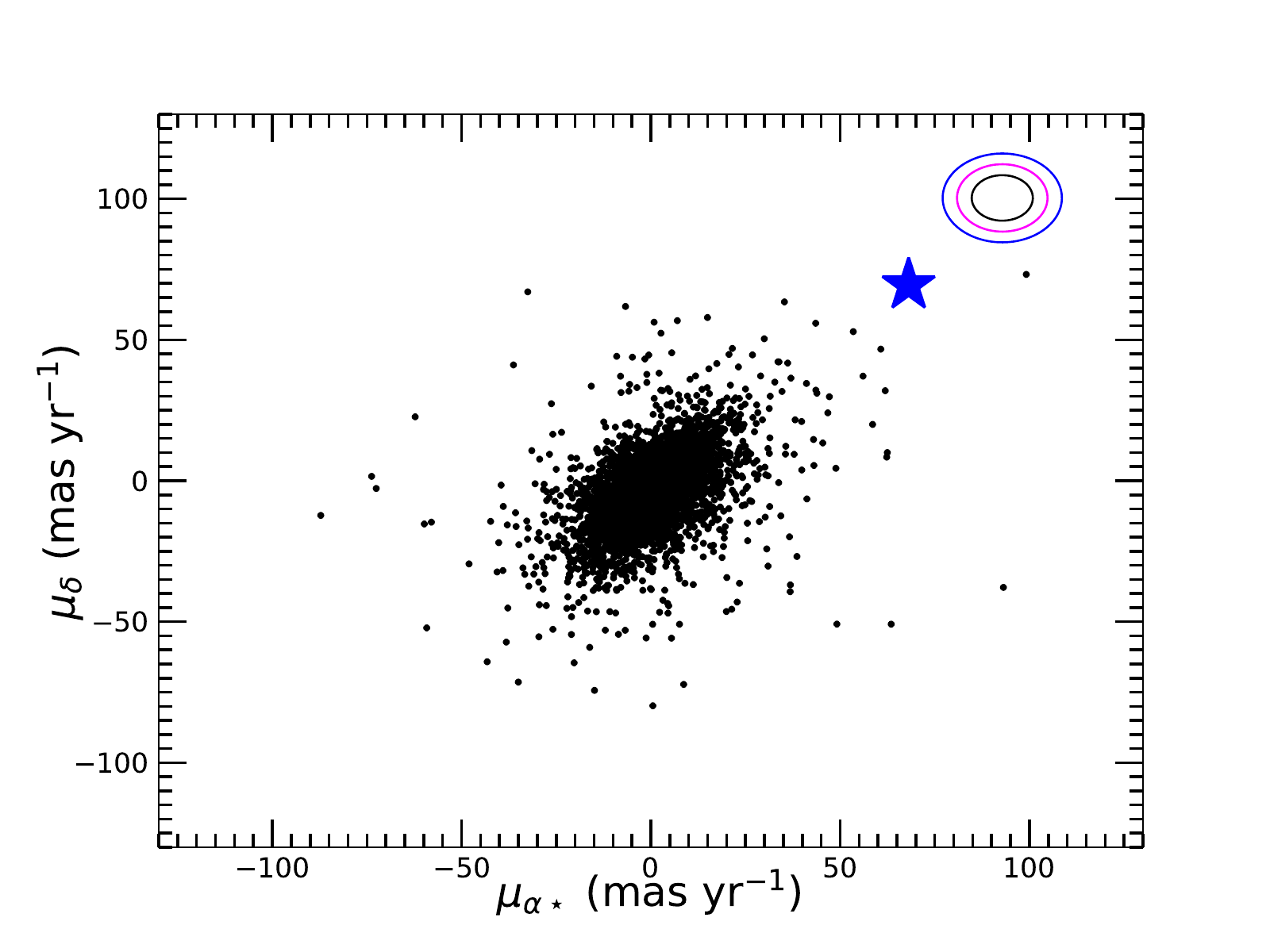} 
   \vspace{-0.1in}
   \captionsetup{labelformat=empty}
  \caption{Figure S7:  \textbf{Proper motion analysis for HIP 99770 b} \cite{Nielsen2017}.   The black dots are simulated proper motions for a population of 10,732 stars synthesized from the Besancon Model of the Galaxy \cite{Robin2003} compared to the 1, 3, and 5-$\sigma$ contours for proper motions needed to match HIP 99770 b's motion across the sky (black, magenta, and blue contours).   The star represents HIP 99770 A's proper motion.  No synthesized star matches HIP 99770 b's proper motion within 5-$\sigma$.}
    \vspace{-0.1in}
    \label{fig:nielsen}
\end{figure*} 

  \begin{figure*}
    \includegraphics[width=1.0\textwidth]{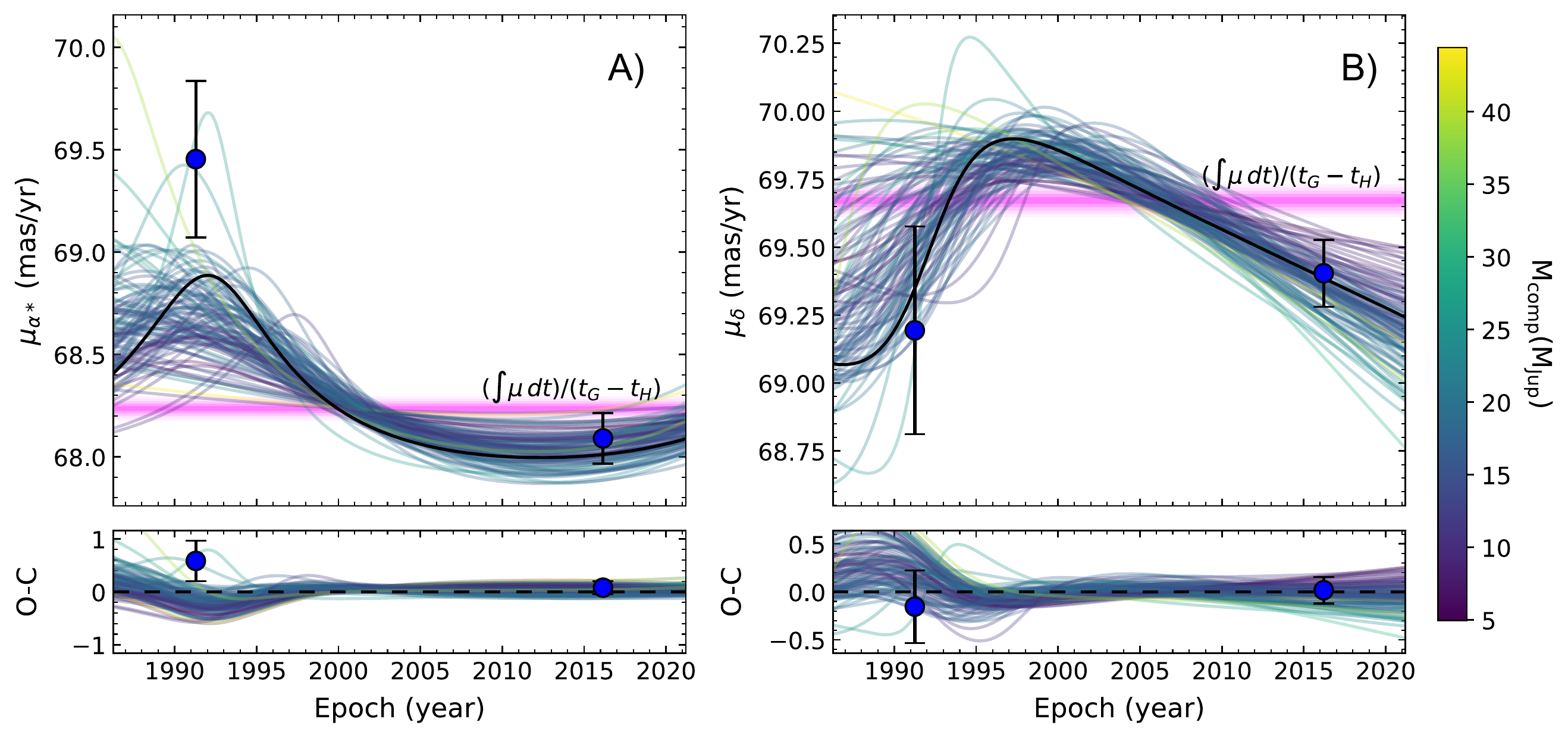} 
   \vspace{-0.3in}
   \captionsetup{labelformat=empty}
  \caption{Figure S8:
   \textbf{Comparison between astrometric motion of the HIP 99770 primary and predicted astrometric motion}.  We display predicted astrometric motion in right ascension (A) and declination (B).  for a sample of 100 orbits drawn from our posterior probability distribution.  The solid black line refers to the best-fitting orbit.  The magenta line shows the average proper motion between Hipparcos and Gaia: the with progressively lighter shadings from the center of this line indicate the 1, 2, 3, 4, and 5-$\sigma$ uncertainties.  The parameter O--C refers to the residuals between the best-fitting orbit and all others.
  }
    \vspace{-0.5in}
\end{figure*} 

  \begin{figure*}
     \includegraphics[width=0.975\textwidth]{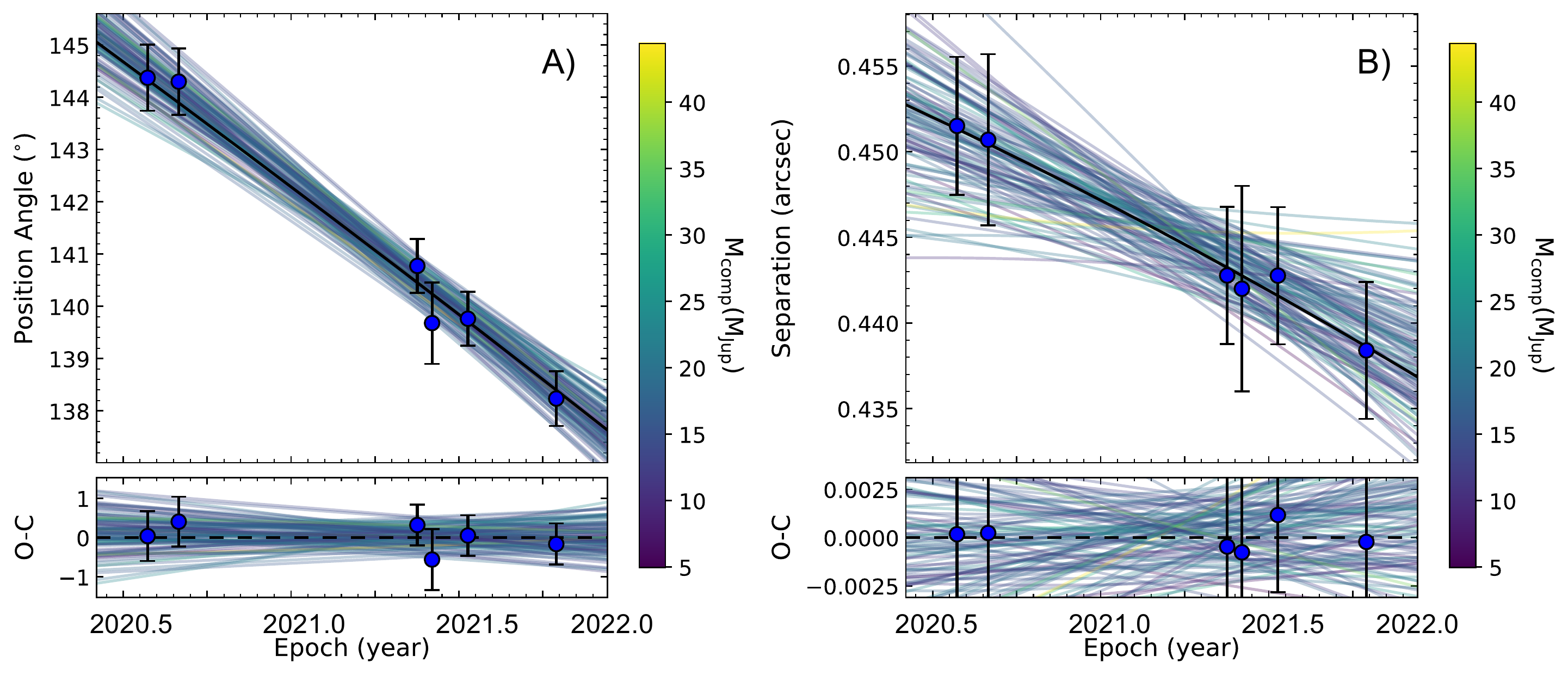}
   \vspace{-0.1in}
   \captionsetup{labelformat=empty}
  \caption{Figure S9:
  \textbf{Comparison between relative astrometry for HIP 99770 b and predicted astrometry}.  We display observed astrometry of HIP 99770 b (data points with 1-$\sigma$ errors) in position angle (A) and angular separation (B).  We display predicted astrometry for the same sample of 100 orbits drawn from our posterior distribution in Figure S8.
  }
    \vspace{-0.5in}
\end{figure*}

 \begin{figure*}[h!]
  \centering
     \includegraphics[width=0.89\textwidth,trim=1mm 1mm 1mm 1mm,clip]{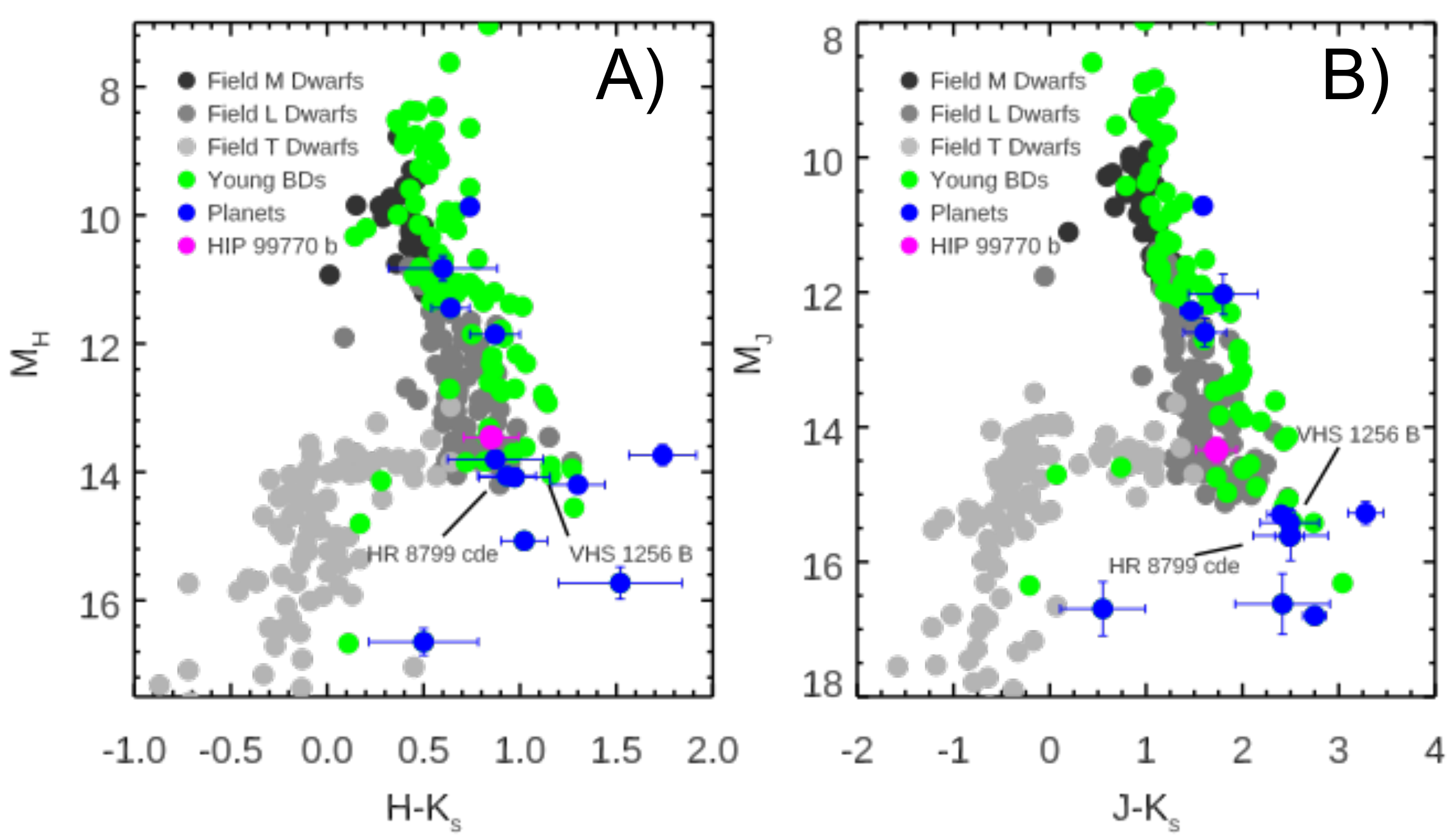} 
   \vspace{-0.1in}
   \captionsetup{labelformat=empty}
  \caption{Figure S10: \textbf{Color-magnitude diagram constraints on HIP 99770 b's atmosphere.} (A) H/H-K$_{\rm s}$ and (B) J/J-K$_{\rm s}$ diagrams comparing HIP 99770 b's photometric properties to field/young brown dwarfs and directly imaged planets.   Data are from Dupuy and Liu \cite{Dupuy2012}. We note the position for the HR 8799 cde planets and the VHS 1256-1257 B substellar companion (19.5 $\pm$ 5 $M_{\rm J}$; 150--300 $Myr$ old) \cite{Marois2008,Marois2010,Gauza2015}.  For VHS 1256-1257 B, we adopt an updated parallax from Dupuy et al \cite{Dupuy2020}.  }
    \vspace{-0.2in}
    \label{fig:empcomp}
\end{figure*} 

 \begin{figure*}[h!]
  \centering
   \includegraphics[width=0.72\textwidth,trim=20mm 0mm 12mm 0mm,clip]{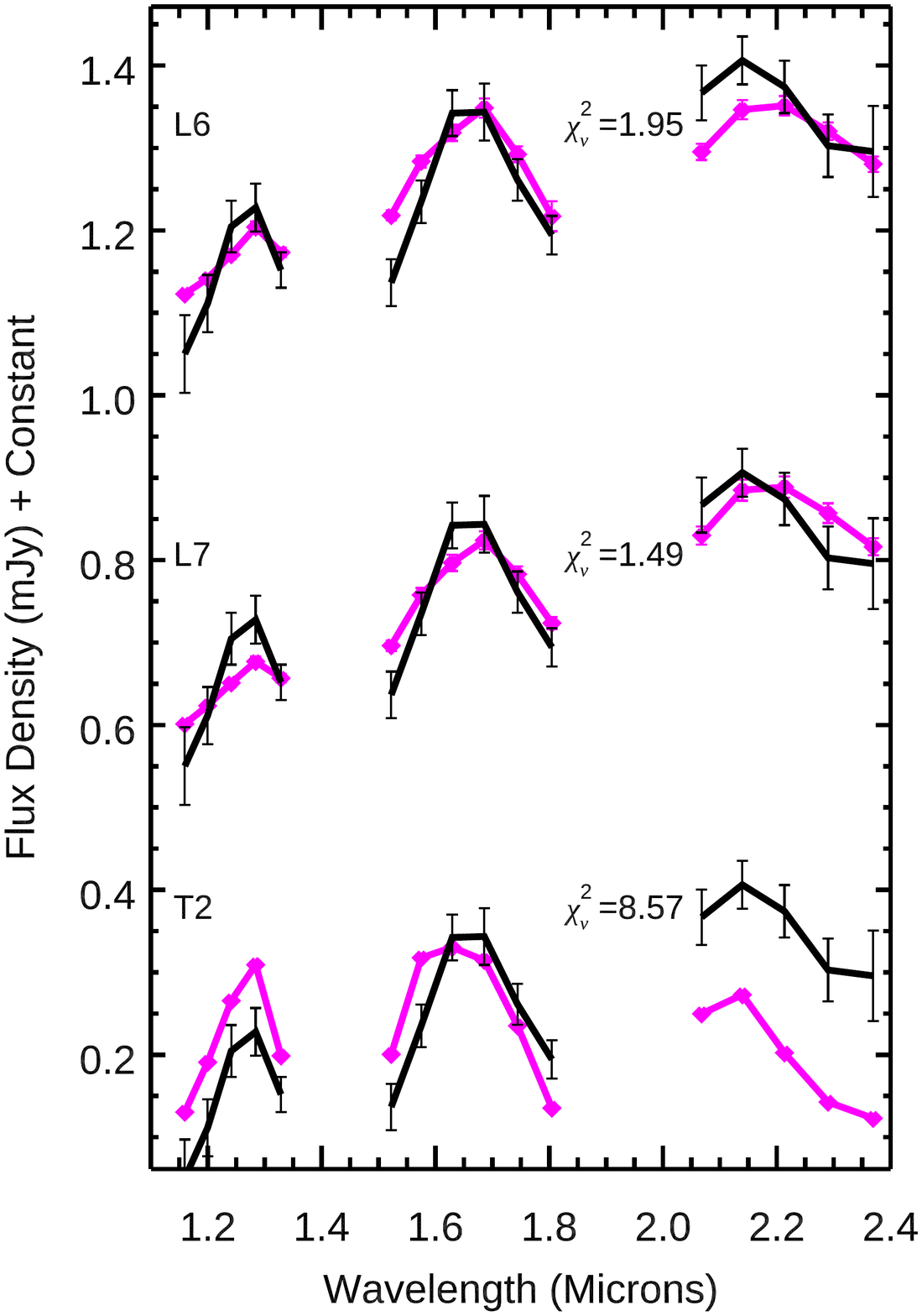} 
   \vspace{-0.1in}
   \captionsetup{labelformat=empty}
  \caption{Figure S11: \textbf{HIP 99770 b's spectrum compared to spectral templates} \cite{Cruz2018}.  HIP 99770 b's spectrum is shown in black with a constant vertical offset.   Spectral templates are shown in magenta. }
    \vspace{-0.2in}
    \label{fig:empcomp}
\end{figure*} 

 \begin{figure*}[h!]
  \centering
  \includegraphics[width=0.95\textwidth,trim=28mm 0mm 12mm 0mm,clip]{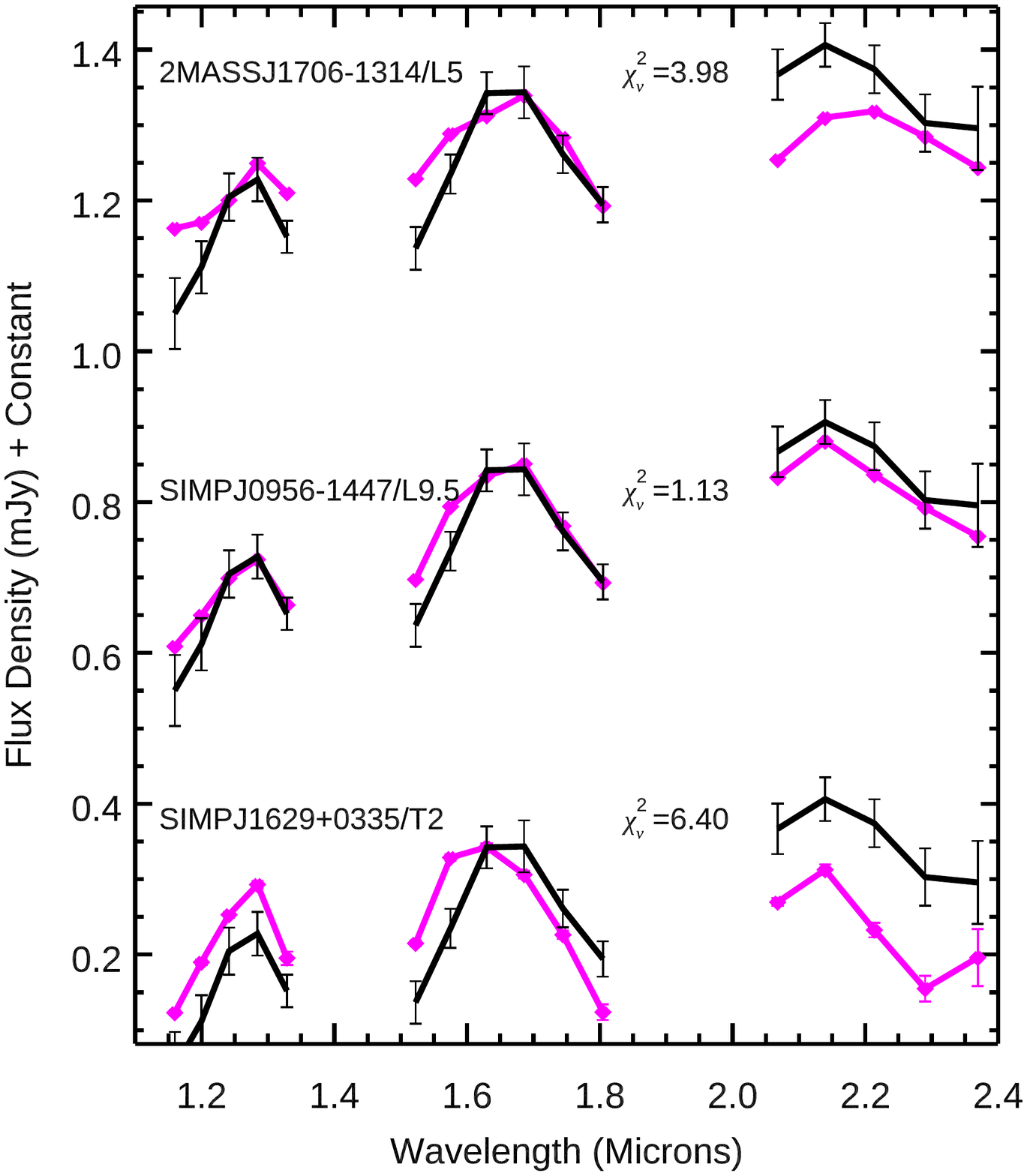} 
   \vspace{-0.3in}
   \captionsetup{labelformat=empty}
  \caption{Figure S12: \textbf{HIP 99770 b's spectrum (black line) compared to an L5, L9.5, and T2 dwarf drawn from the Montreal Spectral Library.} \cite{Gagne2014}.  HIP 99770 b's spectrum is shown in black with a constant vertical offset.   Library spectra are shown in magenta.  }
    \vspace{-0.2in}
    \label{fig:mtllibrary}
\end{figure*} 

\begin{figure*}[h!]
  \centering
  \includegraphics[width=0.995\textwidth,clip]{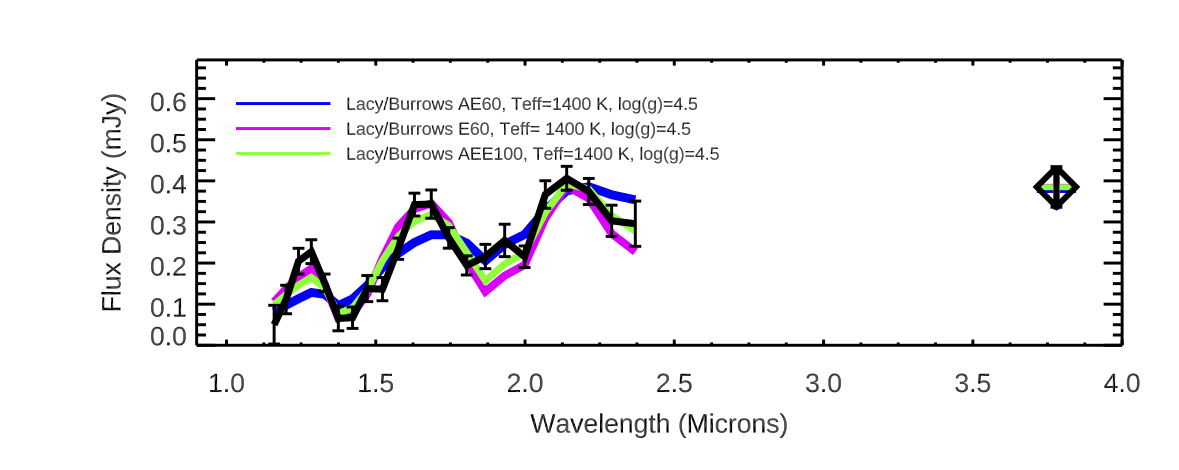} 
   \vspace{-0.4in}
   \captionsetup{labelformat=empty}
  \caption{Figure S13: \textbf{Constraining HIP 99770 b's cloud properties with models.}  Same as Fig 3B, but comparing HIP 99770 b to Lacy/Burrows grid models with different cloud thicknesses/modal dust sizes at a fixed temperature and gravity.   Model grids with very thick clouds needed to reproduce the HR 8799 planet photometry and spectra (AE60) provide a poorer fit than models with thinner clouds and/or larger dust grains (E60, AEE100).   
  }
    \vspace{-0.2in}
    \label{fig:atmosmodels_clouds}
\end{figure*} 

  \begin{figure*}[h!]
  \centering
  \includegraphics[width=0.95\textwidth,trim=0mm 0mm 0mm 0mm,clip]{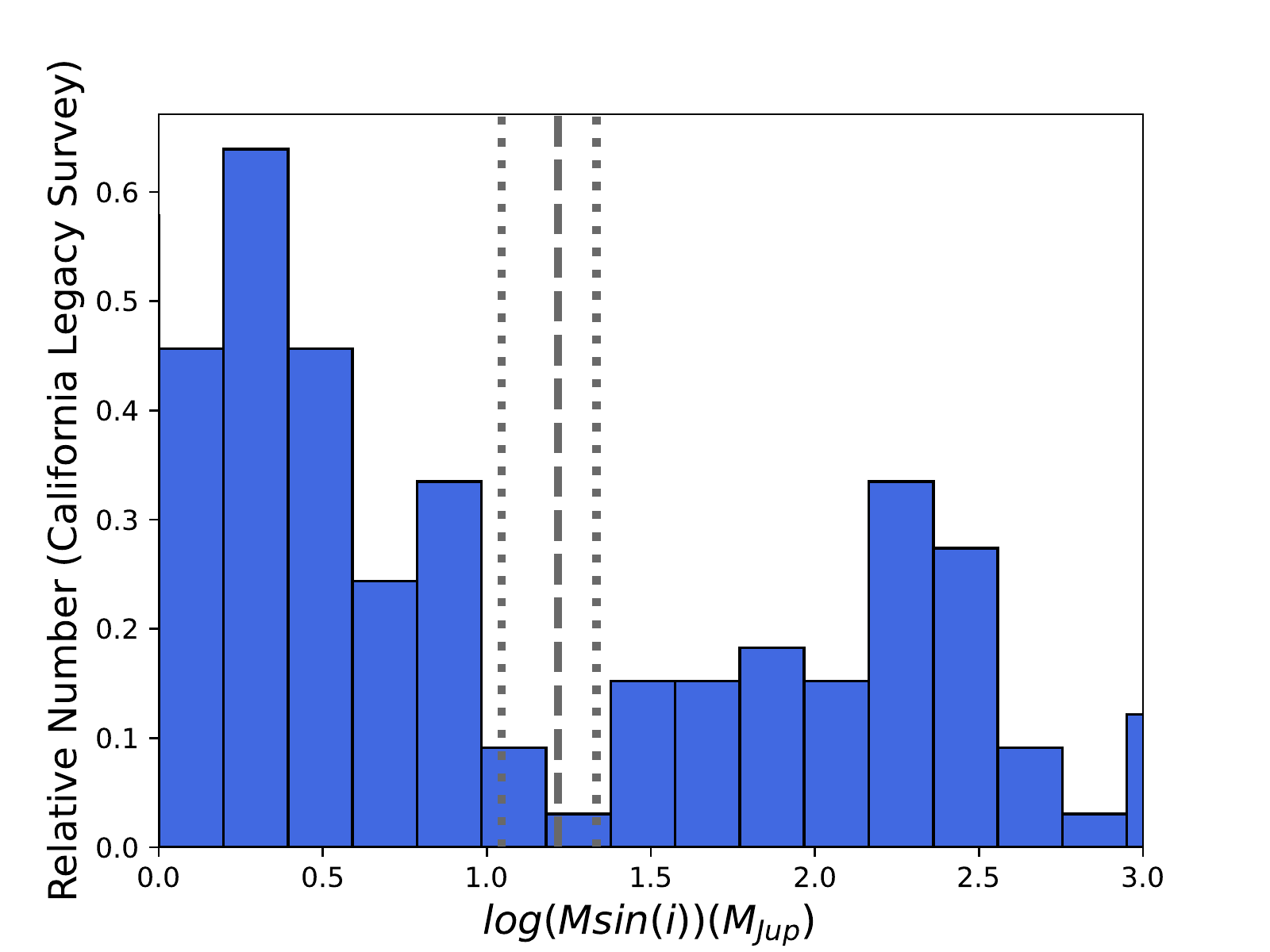} 
   \vspace{-0.0in}
   \captionsetup{labelformat=empty}
  \caption{Figure S14: \textbf{HIP 99770 b's dynamical mass compared to the substellar mass function.}  Histogram plot of the minimum mass ($M~sin(i)$) distribution from the California Legacy Survey for objects orbiting exterior to 0.1 au (i.e. those unaffected by irradiation and tidal disruption).   The minimum occurs at $M~sin(i)$ $\sim$ 16--25 $M_{\rm Jup}$.  Considering inclination effects, the turnover in the companion mass function (nominally separating planets from brown dwarfs) occurs at masses higher than that of HIP 99770 b (vertical dashed line; dashed lines correspond to 1-$\sigma$ range).}
    \vspace{-0.2in}
    \label{fig:bddesert}
\end{figure*} 


\clearpage
\newpage
\textbf{Caption for Data S1:Spectrum of HIP 99770 b extracted from October 2021 data. }. The columns are: 1) Wavelength (microns), 2) Flux Density (in millijanskys), 3) uncertainty in the Flux Density (in millijanskys), and 4) Signal-to-Noise Ratio of the Detection.
\\

\textbf{Caption for Data S2:Lacy \& Burrows atmospheric models.}.  The model filename format is as follows: T[temperature]$\_$g[log of the surface gravity]$\_$[cloud type][modal dust particle size]$\_$[optional flag for reduced methane abundance].21.  E.g. the model T1300$\_$g4.00$\_$AE100.21 corresponds to a temperature of 1300 K, a surface gravity of log(g) = 4, an AE cloud type, and a 100 micron modal dust particle size.  

For each file, the columns are: 1) Grid point number 2) Frequency (Hz) 3) Wavelength (microns) 4) Surface Flux Density, $F_{\rm \nu}$ (ergs/s/cm$^{2}$/Hz) 5) Surface Flux Density, $F_{\rm \lambda}$ (ergs/s/cm$^{2}$/A) 6) Flux Density at 10 pc assuming radii from \cite{Burrows1997,Burrows2006} (millijanskys) 7) Wavelength (angstroms) 8) Smoothed Flux Density at 10 pc assuming radii from \cite{Burrows1997,Burrows2006} (millijanskys) 9) Brightness Temperature (at the photosphere, tau = 2/3) 10) Smoothed Brightness Temperature (at the photosphere, $\tau$ = 2/3) 11) Smoothed Surface Flux Density, $F_{\rm \lambda}$ (ergs/s/cm$^{2}$/A) 12) Pressure at the photosphere (where $\tau$ = 2/3, in units of bars).

\end{document}